\documentclass[10pt, letterpaper, conference]{IEEEtran}
\IEEEoverridecommandlockouts

\usepackage[table]{xcolor}
\usepackage{annotations}
\usepackage{enumitem}

\usepackage{pifont}
\definecolor{cadmiumgreen}{rgb}{0.0, 0.42, 0.24}
\usepackage{multirow}
\usepackage{setspace} 
\usepackage{soul}
\usepackage{mathtools}
\usepackage{amsmath,amssymb,amsthm}
\newcommand{\solution}{ALMOST}

\def\smallerspacecaption{\vspace{-0.5em}}
\def\smallerspacebelowfigure{\vspace{-1.5em}}

\usepackage{setspace}
\usepackage{nicefrac}
\usepackage{siunitx}
\usepackage{array,framed}
\usepackage{booktabs}
\usepackage{
  color,
  float,
  epsfig,
  wrapfig,
  graphics,
  graphicx,
}
\usepackage[caption=false,font=footnotesize]{subfig} 

\usepackage{setspace}
\usepackage{amsfonts}
\usepackage{latexsym,fancyhdr,url}
\usepackage[linesnumbered,ruled,vlined]{algorithm2e}
\usepackage{algpseudocode}
\usepackage{graphics}
\usepackage{xparse} 
\usepackage{xspace}
\usepackage{multirow}
\usepackage{csvsimple}
\usepackage{balance}

\usepackage{mathtools}

\DeclareMathAlphabet{\mathcal}{OMS}{cmsy}{m}{n}

\usepackage{multirow}
\usepackage{booktabs}
\usepackage{tabularx}
\newcolumntype{L}[1]{>{\raggedright\let\newline\\\arraybackslash\hspace{0pt}}m{#1}}
\newcolumntype{C}[1]{>{\centering\let\newline\\\arraybackslash\hspace{0pt}}m{#1}}
\newcolumntype{R}[1]{>{\raggedleft\let\newline\\\arraybackslash\hspace{0pt}}m{#1}}

\newcommand{\blue}[1]{\textcolor{blue}{#1}}

\usepackage{etoolbox}
\patchcmd{\quote}{\rightmargin}{\leftmargin 2em \rightmargin}{}{}

\usepackage{listings}

\newcommand{\listingcaption}[1]%
{%
\refstepcounter{lstlisting}{\hspace{-15pt} Listing \thelstlisting: #1}
}%
\newlength{\MaxSizeOfLineNumbers}%
\usepackage[ruled]{algorithm2e}

\SetAlCapNameFnt{\footnotesize}
\SetAlCapFnt{\footnotesize}
\SetAlCapHSkip{0pt}
\IncMargin{-\parindent}
\usepackage{algorithmicx}

\usepackage{enumitem}

\usepackage{acronym}
\acrodef{IC}{integrated circuit}
\acrodef{EDA}{electronic design automation}
\acrodef{HDL}{hardware description language}
\acrodef{AIG}{and-inverter-graph}
\acrodef{ML}{machine learning}
\acrodef{IP}{intellectual property}
\acrodef{RTL}{register transfer level}
\acrodef{DAG}{directed acyclic graph}
\acrodef{GCN}{graph convolutional network}
\acrodef{SoC}{system-on-chip}
\acrodef{DNN}{deep neural network}
\acrodef{MDP}{Markov decision process}
\acrodef{CNN}{convolutional neural network}
\acrodef{MSE}{mean-square error}
\acrodef{SA}{simulated annealing}
\acrodef{SOTA}{state-of-the-art}
\acrodef{GNN}{graph neural network}
\acrodef{KPA}{key prediction accuracy}
\acrodef{OMLA}{oracle-less machine learning attack}
\acrodef{PPA}{power, performance, and area}
\acrodef{Acc}{accuracy}
\acrodef{wolog}{without loss of generality}

\usepackage{booktabs}    
\usepackage{tabularx}
\usepackage{booktabs}    
\usepackage{multirow}
\usepackage{graphicx}

\usepackage{xparse}
\newcommand{\bnm}{\begin{newmath}}
\newcommand{\enm}{\end{newmath}}

\newcommand{\bea}{\begin{eqnarray*}}%
\newcommand{\eea}{\end{eqnarray*}}%

\newcommand{\bne}{\begin{newequation}}
\newcommand{\ene}{\end{newequation}}

\newcommand{\bal}{\begin{newalign}}
\newcommand{\eal}{\end{newalign}}

\newenvironment{newalign}{\begin{align}%
\setlength{\abovedisplayskip}{4pt}%
\setlength{\belowdisplayskip}{4pt}%
\setlength{\abovedisplayshortskip}{6pt}%
\setlength{\belowdisplayshortskip}{6pt} }{\end{align}}

\newenvironment{newmath}{\begin{displaymath}%
\setlength{\abovedisplayskip}{4pt}%
\setlength{\belowdisplayskip}{4pt}%
\setlength{\abovedisplayshortskip}{6pt}%
\setlength{\belowdisplayshortskip}{6pt} }{\end{displaymath}}

\newenvironment{newequation}{\begin{equation}%
\setlength{\abovedisplayskip}{4pt}%
\setlength{\belowdisplayskip}{4pt}%
\setlength{\abovedisplayshortskip}{6pt}%
\setlength{\belowdisplayshortskip}{6pt} }{\end{equation}}

\newcounter{ctr}

\newcounter{mytable}
\def\mytable{\begin{centering}\refstepcounter{mytable}}
\def\endmytable{\end{centering}}

\newcounter{myfig}
\def\myfig{\begin{centering}\refstepcounter{myfig}}
\def\endmyfig{\end{centering}}

\newlength{\saveparindent}
\newlength{\saveparskip}
\newcommand{\E}{{\rm I\kern-.3em E}}

\renewcommand{\eqref}[1]{\mbox{Equation~(\ref{#1})}}

\def \part {part}

\DeclareMathOperator*{\argmin}{argmin}

\renewcommand{\paragraph}[1]{\vspace*{6pt}\noindent\textbf{#1}\;}

\def \blackslug{\hbox{\hskip 1pt \vrule width 4pt height 8pt
    depth 1.5pt \hskip 1pt}}
\def \qed{\quad\blackslug\lower 8.5pt\null\par}

\newcounter{mynote}[section]

\newcommand\ignore[1]{}

\newcounter{rcnote}[section]

\newcounter{mrnote}[section]

\newcounter{fknote}[section]

\newcounter{anote}[section]

\DeclareMathSymbol{\mlq}{\mathord}{operators}{``}
\DeclareMathSymbol{\mrq}{\mathord}{operators}{`'}

\newcommand{\rhf}[2]{R_{f, \gamma}}

\DeclareDocumentCommand{\edist}{o o}{
  \ensuremath{
    \IfNoValueTF{#1}{{d}}{{\sf d}(#1,#2)}
  }
}

\newcommand{\olrk}[1]{\ifx\nursymbol#1\else\!\!\mskip4.5mu plus 0.5mu\left(\mskip0.5mu plus0.5mu #1\mskip1.5mu plus0.5mu \right)\fi}

\NewDocumentCommand{\indseq}{ O{1} O{r} }{{#1}\ldots {#2}}

\newcommand{\circleone}{\ding{202}}
\newcommand{\circletwo}{\ding{203}}

\begin{document}

\title{ALMOST: \underline{A}dversarial \underline{L}earning to \underline{M}itigate \underline{O}racle-less ML Attacks via \underline{S}ynthesis \underline{T}uning}

\author{\IEEEauthorblockN{Animesh B. Chowdhury\IEEEauthorrefmark{1}, Lilas Alrahis\IEEEauthorrefmark{2}, Luca Collini\IEEEauthorrefmark{1}, Johann Knechtel\IEEEauthorrefmark{2}, 	\\Ramesh Karri\IEEEauthorrefmark{1}, Siddharth Garg\IEEEauthorrefmark{1}, Ozgur Sinanoglu\IEEEauthorrefmark{2}, Benjamin Tan\IEEEauthorrefmark{3}}
\IEEEauthorblockA{\IEEEauthorrefmark{1}\textit{New York University, USA}, \IEEEauthorrefmark{2}\textit{New York University Abu Dhabi, UAE}, \IEEEauthorrefmark{3}\textit{University of Calgary, Canada}}
}

\date{November 2019}
\maketitle

\IEEEpubidadjcol

\begin{abstract}
%
Oracle-less machine learning (ML) attacks have broken various logic locking schemes.
Regular synthesis, which is tailored for area-power-delay optimization, yields netlists where key-gate localities 
are vulnerable to learning.
Thus, we call for security-aware logic synthesis.
We propose ALMOST, a framework for adversarial learning to mitigate oracle-less ML attacks via synthesis tuning.
ALMOST uses a simulated-annealing-based synthesis recipe generator, employing adversarially trained models that
can predict state-of-the-art attacks' accuracies over wide ranges of recipes and key-gate localities.
Experiments on ISCAS benchmarks confirm the attacks' accuracies drops to around 50\% for ALMOST-synthesized
circuits, all while not undermining design optimization.

\end{abstract}


\section{Introduction}

\textbf{Setting:}
\Ac{ML}-based structural attacks like~\cite{omla,SAIL} can decipher the key-bits of logic-locked circuits without access to an
\textit{oracle}.\footnote{An oracle provides black-box access to the correct functionality, e.g., through a functional chip
	obtained from the open market.}
Such attacks explored and exploited the impact of logic synthesis on structural properties of locked
netlists.
For example, even with bubble pushing -- which is a
logic synthesis technique used by locking schemes to obfuscate the relationship between key-bits to key-gates -- some
discernible structures can arise.
This is because, even though logic synthesis is complex and uses carefully devised \ac{PPA} heuristics, it is deterministic, yielding predictable structural transformations around key-gates, which can be learned on to infer key-bits.
%



\textbf{Prior Art:}
Truly random logic
locking (TRLL)~\cite{limaye2020thwarting} and UNSAIL~\cite{alrahis2021unsail} are assumed to
thwart ML attacks -- both schemes remain unbroken as of today. However, both schemes also rely on specific design
properties, like certain structures of gates to be present in sufficient numbers in the appropriate places, restricting
their general applicability.
Both schemes operate on netlists handled by commercial synthesis tool, which are operating as block-box
modules that optimize for \ac{PPA}, not for security.
Thus, both schemes needed to ``work around'' those tools in a smart way to enable resilient locking schemes.


\textbf{Challenge:}
There is a need for logic-locking techniques that are a) resilient against ML-based, oracle-less attacks
exploiting structural properties of locked netlists, b) generally applicable to various designs, and c) integrate with
logic synthesis and do not undermine design optimization offered by synthesis.


%

\begin{figure}[t]
    \centering
    \includegraphics[width=0.85\columnwidth]{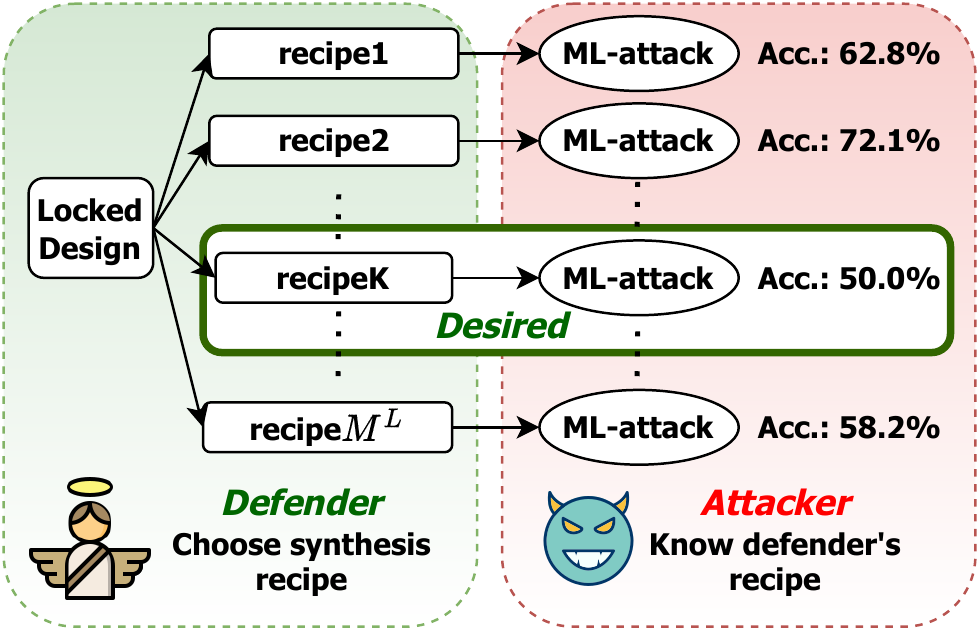}
\smallerspacecaption
    \caption{Different synthesis recipes have different impacts on the resilience of a logic-locked design in the context
	    of ML-based attacks.}
    \label{fig:almostMotivation}
\smallerspacebelowfigure
\end{figure}

\textbf{Scope:} We study the causal nexus between logic locking, logic synthesis, and resilience to ML-based attacks. 

For the first time, we propose a logic-locking methodology that drives logic synthesis to makes the locked circuits resilient
against ML-based attacks.
Unlike TRLL or UNSAIL, we achieve resilient
against ML-based attacks without dedicated locking strategies or crafted key-gate structures.
We demonstrate that we can use the simplest and vulnerable locking
scheme, i.e., random logic locking (RLL)~\cite{epic}, and still obtain ML-resilient designs.

However, achieving theis resilience requires tackling practical challenges.
As shown in Fig.~\ref{fig:almostMotivation}, even when starting from the same locked design, varying synthesis recipes
will impact the resiliency against ML attacks differently. Finding synthesis recipes optimized for a specific metric,
e.g., the accuracy of ML attacks\footnote{%
Accuracy is an established ML metric. It is defined as
(\# {correctly predicted key-bits}) / ({total} \# {key-bits}, i.e., key-size). For a key-size
of 128 bits, when an attack correctly predicts 64-of-the-128  bits (and the remaining 64 bits are either incorrectly
		inferred or not inferred at all), the accuracy is 0.5 or 50\%. As a defender, accuracy
 of 50\% is desired. This equates to the attack is  no better  than a random
guess.},
is $\Sigma^2_p$-hard~\cite{umans1999hardness}.
Thus, our study addresses two key research challenges (RCs).
\begin{enumerate}
 \item\textbf{RC1:} Can we explore the synthesis search space to find some recipe(s) that
 provide resilience against various \ac{SOTA} ML attacks?
 \item\textbf{RC2:} Given any recipe, can we efficiently quantify its \ac{ML} resiliency without 
 running the various \ac{SOTA} attacks?
\end{enumerate}

\textbf{Contributions:}
\solution{} is a security-centric framework to generate ML-resilient logic-locked designs using synthesis. It does not relying on security promises of the locking scheme. This is important as the value of such promises are subject to synthesis in the first place.
%
%
Our contributions include:












\begin{enumerate}
    \item A framework for {security-aware synthesis-recipe generation} using \ac{SA}. These recipes produce
    netlists that are ML attack resilient (RC1).
    \item An adversarially trained model to predict resiliency of locked designs post-synthesis
    against ML attacks
    (RC2).
    \item Public release of source codes and artifacts. 
\end{enumerate}

 \textbf{Key Results:} For \solution{}-synthesized ISCAS85 benchmarks, using RLL as an
 exemplary scheme that is otherwise fully vulnerable, \ac{SOTA} ML attacks~\cite{omla,alaql2021scope} {and non-ML structural attacks~\cite{redundancy}}
 are reduced to random guessing.
It is important to note that \solution{} generates synthesis recipes that enforce/maintain such resiliency even when the
ML-empowered attacker is fully aware of the recipe.
 At the same time, PPA overheads are marginal.






\section{Oracle-Less Attacks on Logic Locking}
\label{sec:background}

Recent works show the potential of ML models for advanced attacks in the oracle-less scenario~\cite{SAIL, snapshot, omla, alaql2021scope}.

\textit{SAIL}~\cite{SAIL} tackles XOR/XNOR locking. It learns the local, synthesis-induced changes around those key-gates,
reverting the binding of key-bits: before bubble pushing, XOR is bound to `0' and XNOR to `1'.
\textit{SnapShot}~\cite{snapshot} works by learning the local changes caused by key insertion
and synthesis itself.

While SAIL and SnapShot use classic tensor-based models, \textit{OMLA}~\cite{omla} uses \acp{GNN}, as 
a natural representation of gate-level netlists.
OMLA extracts {\textit{localities}, i.e., the sub-circuit structures} around key-gates, represents them
as sub-graphs, and passes them to the GNN to
predict the corresponding key-bits through subgraph classification.

These attacks employ {self-referencing} to train their ML models. The attacks apply some form
of re-locking and re-synthesis to generate training datasets. The attacks know the synthesis recipe used by the
defender.

\textit{SCOPE}~\cite{alaql2021scope} extracts synthesis reports, e.g., on area and power, to decipher key-bits.
Unlike the attacks above, SCOPE follows an unsupervised approach for learning the correlation between key-bit
values and synthesis features.

\textit{Redundancy attack}~\cite{redundancy} is a non-ML structural attack. It assumes that the original design is
fully testable. Accordingly, the attack infers the key-bits as those assignments that
cause fewer untestable faults in the locked circuit.

\section{\solution{} framework}

We formulate security-aware synthesis as an optimization problem where the objective is to ensure that locked designs
remain resilient post-synthesis.
Thus, we search for synthesis recipes that structure the
netlists such that \ac{SOTA} attacks, be they ML or non-ML ones, can achieve only $\sim$50\% accuracy.

\Ac{wolog}, we use the open-source synthesis suite \textit{yosys} and \textit{ABC}~\cite{brayton2010abc}. Unlike
commercial tools, this suite allows for fine-grain tuning of synthesis recipes, also accounting for user-defined
objectives as needed, i.e., for accuracy in this work.
%
Like any synthesis tool, this suite takes \ac{RTL} and converts it to a gate-level netlist, by performing
technology mapping using a technology library. Synthesis transformation and optimization steps are
implemented using the \ac{AIG} representation.

\begin{figure}[t]
    \centering
    \includegraphics[width=\columnwidth]{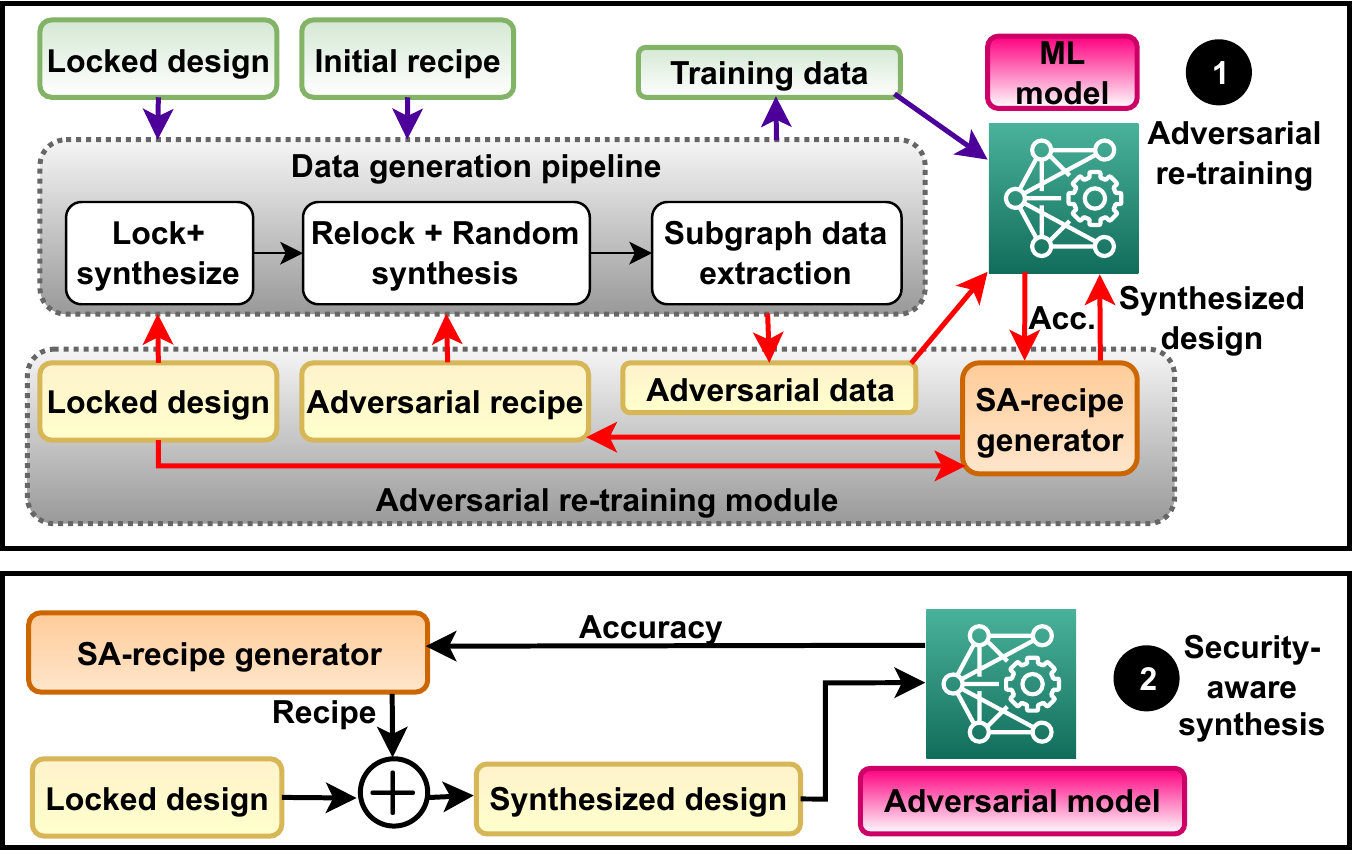}
    \smallerspacecaption
    \smallerspacecaption
    \smallerspacecaption
    \smallerspacecaption
    \caption{ALMOST framework}
    \label{fig:almostFramework}
\smallerspacebelowfigure
\end{figure}

Next, we formulate the problem in detail. Then, we discuss the framework stages, which
are also illustrated in Fig.~\ref{fig:almostFramework}.



\subsection{Problem Formulation}
\label{sec:problem_form}

\textbf{General Problem.}
In a synthesis recipe $\mathbf{S}$ with $\mathbf{M}$ transformation steps, the number of recipes of length $\mathbf{L}$ is $\mathbf{M}^\mathbf{L}$. Security-aware recipe generation can be formulated as:
\begin{gather}
\argmin_{\mathbf{S}} |Acc_{M_A^S}(\mathcal{G} (\mathbf{AIG},\mathbf{S}))-0.5|
\label{eq:omlaopt}
\end{gather}
where $\mathcal{G}$ is the synthesis function defined as $\mathcal{G} : \mathbf{AIG} \times \mathbf{S} \longrightarrow
\mathbf{AIG}$. \Ac{Acc} describes
the prediction accuracy of an attacker's model $M_A^S$
that is built-up using recipe $\mathbf{S}$.
\Ac{Acc} evaluates the {resilience of the locked netlist introduced through} synthesis using $\mathbf{S}$; \ac{Acc}
values around 50\% are the target.

Note that we do not explicitly consider PPA in the above formulation, but only security. Nevertheless, we observe
empirically that (i) PPA optimization can follow-up on our security-aware synthesis without undermining the netlists'
resilience and (b) PPA overheads are, on average, only marginal.

\textbf{Model $M_A^S$ and Naive Approach.}
We aim to solve the optimization problem (Eq.~\ref{eq:omlaopt}) using a black-box solver, \ac{wolog} \ac{SA} in this
work.
The challenge here is that, to accurately evaluate the effects of the attacker's model $M_A^S$,\footnote{%
Typically, $M_A^S$ is a binary classifier trained to minimize the loss function
\begin{gather}
    \hat{\theta} = \argmin_\theta \frac{1}{n} \sum_{i=1}^n L(M_A^S({\theta};x_{i}), y_i) := \mathcal{L}(X;Y,\theta).
\end{gather}
where $\{x_i, y_i\}$, $i=1,\ldots,n$ are entires of labeled dataset $X$, $Y$ and $\hat{w}$ is a trainable parameter.
For ML attacks, $X$ denotes the feature embeddings of key-gates of a relocked and resynthesized design, and $Y$
denotes their corresponding key-bit values. $n$ denotes the total number of locations relocked in the design.
The procedure for labeled dataset generation involves re-locking the locked netlist under attack and then 
re-synthesizing it using the defender's recipe $S$
This approach is taken by the SOTA ML attacks
(Sec.~\ref{sec:background}); it is
is based on the insight that doing so
allows to accurately capture the structural transformations caused by $S$.
}
we would need to separately train models $M_A^S$ for the varying recipes $S$ across every iteration.
Fig.~\ref{fig:almostMotivation} outlines such naive approach,
which seems
computationally expansive.

\textbf{Challenge for Transferability of $M_A^S$.}
While $M_A^S$ models are demonstrated to predict key-bits very well, their accuracy would drop for locked designs that are
synthesized using any other recipe, say $S'$, i.e., accuracy($M_A^S$) $\leq$ accuracy($M_A^{S'}$).
This is because the range of structural transformations induced by $S$ may not fully match with the range induced by
$S'$.

To confirm this intuition, we run an experiment on the ISCAS85 circuit c5315 where we trained two attack models,
$M_A^{S_1}$ and $M_A^{S_2}$, with training data covering the locked netlist $T_{S_1}$ and $T_{S_2}$ as synthesized using recipes $S_1$ and
$S_2$, respectively.
%
We observe that accuracy($T_{S_1}$,$M_A^{S_1}$) = 57.52\%, whereas accuracy($T_{S_1}$,$M_A^{S_2}$) = 52.27\%.
Similarly,  accuracy($T_{S_2}$,$M_A^{S_1}$) = 53.78\% and accuracy($T_{S_2}$,$M_A^{S_2}$) = 58.91\%.
(We report the setup and more detailed results in Sec.~\ref{sec:comparingOMLAmodels}.)

These accuracy mismatches clearly show challenges for transferring $M_A^S$ for the evaluation of netlists
locked with other recipes, thus re-iterating the need for training unique models for each iteration of optimization
and exploration of the synthesis search space (Fig.~\ref{fig:almostMotivation}).
However, as indicated, training of separate models for every iteration seem too expensive.

\begin{figure}
    \centering
    \includegraphics[width=\columnwidth]{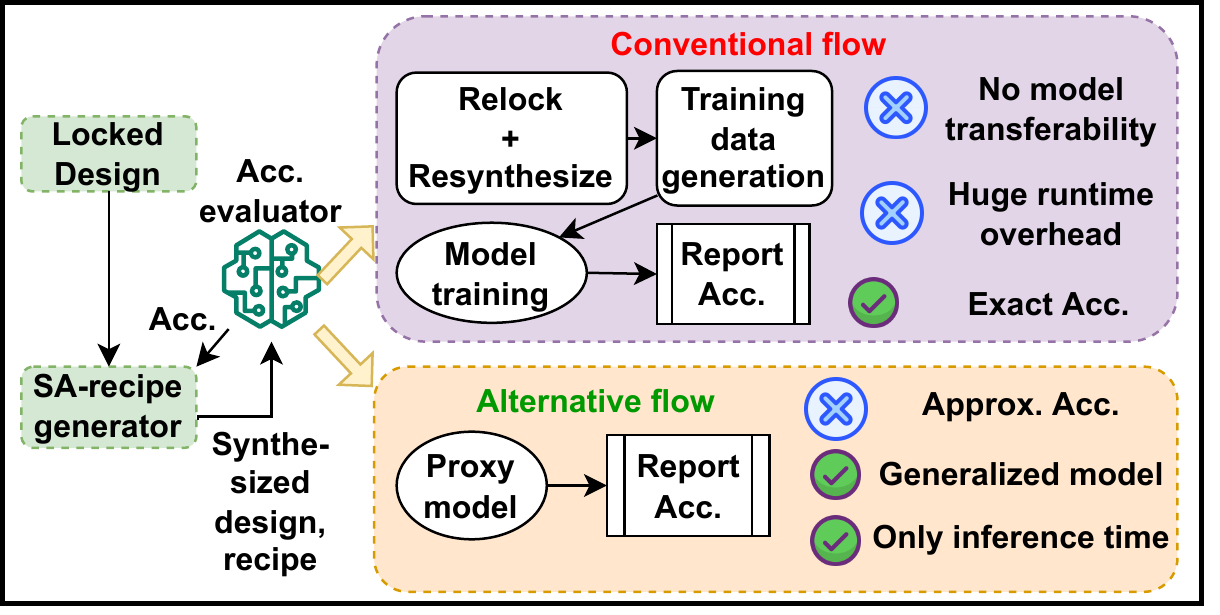}
\smallerspacecaption
\smallerspacecaption
\smallerspacecaption
\smallerspacecaption
    \caption{Motivation for a proxy model.}
    \label{fig:omlaIssue}
\end{figure}

\textbf{Proposed Solution.}
To enable a practical exploration of the search space, i.e., to tackle the stated RCs, we need an
alternative in form of a proxy model $M_A^*$ that yields good estimate of $M_A^S, \forall S \in [1,M^L]$.
The challenge and motivation for such proxy model are also illustrated in Fig.~\ref{fig:omlaIssue}.

For training such high-quality, transferable proxy model,
the training data should contain a good number of structural transformations that are
observed for a range of synthesis recipes.
Next, we discuss how to train such proxy model.

\begin{algorithm}[t]
\caption{Adversarial ML attack model training}
\label{algo:OMLAadv}
\footnotesize
\KwData{Re-synthesized locked netlist: $AIG_{initial}$, Epochs: $N$, Batchsize: $B$, Learning rate: $\gamma$, Periodicity: $R$} 
\KwResult{Adversarial model $M_A^*(\cdot;\theta_{adv})$}
Relock + Resynthesize $AIG_{initial}$ with random $L=10$ length recipes.\\
Create $D_{training}=(X_{train},Y_{train})$ using subgraph extraction from key-gates. \\
$\theta_{0} \leftarrow$ He initialization, $t \leftarrow 0$\\
\While{$t$ $< N$}{
            \If{$t \% R$}{
                $s^* \leftarrow$ SA ($T_{init}$,$M_A^*(\cdot;\theta^t),N_{max}$,AIG$_{initial}$) \\
                Compute $X^{s^*},Y^{s^*}$ and augment $D_{training}$
            }
			Compute $\Delta \theta^{(t)} = - \nabla \mathcal{L}(\theta^{(t)})$\label{lin:deep-learning-delta-w}\\
			$\theta^{(t + 1)} := \theta^{(t)} + \gamma \Delta \theta^{(t)}$ \\
\Return $M_A^*(\cdot;\theta_{adv})$
}
\end{algorithm}

\subsection{Adversarially Trained Attacker's Model $M_A^*$ (\circleone)}

We adopt adversarial re-training~\cite{madry2017towards,entropysgd}
to train a model that
is exposed to a wide variation of localities, i.e., subgraph structures around key-gates.
The motivation for adversarial retraining is as follows: we want to create adversarial subgraph embeddings where the
attack model $M_A^*$ mispredicts, whereupon we use these adversarial subgraphs to augment the training data and to
subsequently learn
a more robust model.

We create adversarial subgraph embeddings differently to conventional $\ell_2$ or $\ell_{\inf}$ perturbations in the
image/vision domain~\cite{madry2017towards}. Since logic synthesis offers no analytical closed-form function, we
do
gradient-free optimization. Similar to adding $\delta$ perturbations to original images, we seek for a synthesis recipe
$S_{adv}$ which transforms the re-locked netlist such that the subgraph embeddings are misclassified. We solve an optimization problem for obtaining adversarial samples:
\begin{gather}
{x}^{\text{S}} = \arg\max_{\mathbf{S}} L(M_A^*(\theta;\hat{x}), y) \\
\text{where } \hat{x} = \textbf{READOUT}{(h^k_{AIG^S}, k \in K)}, \\
AIG^S\leftarrow \mathcal{G}(AIG,S)
\label{eq:attack}
\end{gather}

To solve this gradient-free optimization problem, we use \acf{SA}. Once the adversarial samples are generated, we
augment the training data, rendering the model more robust and generalizable. Thus, we sample synthesis recipes
that can provide diverse variants of the locked netlist under consideration. We solve the following
min-max objective function for the adversarially trained model $M_A^*$. 
\begin{gather}
 \hat{\theta} = \min_\theta \max_{\mathbf{S}} \frac{1}{m} \sum_{i=1}^{m} L(M_A^*({\theta};x^S_{i}), y_i)
\end{gather}
We outline more details in Algorithm~\ref{algo:OMLAadv}.

\subsection{Black-Box Optimization for Security-Aware Synthesis (\circletwo)}
While \solution{} may use a variety of black-box optimization approaches, like
evolutionary algorithms, tree-search, etc., we use, \ac{wolog}, a standard procedure for \ac{SA}.






\section{Experimental Evaluation}\label{sec:exp}

Next we show
how locked and \solution{}-synthesized designs are resilient
against SOTA attacks, as the attack accuracy approaches random guessing. 

\subsection{Setup}
\textbf{Benchmarks.} We evaluate \solution{} on the largest ISCAS85 combinational benchmarks. We initially lock them
with RLL, considering key-sizes of $64$ and $128$. The resilience of ALMOST is demonstrated against OMLA, SCOPE, and
the redundancy attacks (Sec.~\ref{sec:background}).

\textbf{Synthesis.} As indicated, we use the synthesis suite \textit{yosys} with \textit{ABC}~\cite{brayton2010abc}, for its
flexibility of tuning synthesis recipes.

We select \textit{resyn2} as baseline recipe, which is widely used for delay optimization. For fair comparison,
   we devise fixed-length synthesis recipe of $L=10$. We employ seven synthesis transformations: \texttt{rewrite,
	   re-substitute, refactor, rewrite -z, resub -z, refactor -z}, and \texttt{balance}.
	   For technology mapping, we use the NanGate 45~nm technology library.

\textbf{Attack Model.} As the OMLA framework is publicly available~\cite{omla}, we employ it for building the attacker's models.  
We set the network architecture, training configuration, and hyper-parameter settings as reported in~\cite{omla}.
For additional characterization of \solution{}, we apply recent oracle-less attacks, SCOPE~\cite{alaql2021scope} and Redundancy~\cite{redundancy}.

We study the effectiveness of adversarially
trained $M_A^*$ by comparing three variants:
\begin{itemize}
    \item $M_A^{resyn2}$ is the baseline model, where the attacker re-locks and re-synthesizes using the defender's
    baseline synthesis recipe, resyn2.
    \item $M_A^{random}$ is trained on re-locked circuits that are re-synthesized using random recipes of length
    $L=10$.
    \item $M_A^*$ is trained using our adversarial data-augmentation-based re-training.
\end{itemize}

For \textbf{adversarial training}, we generate adversarial samples after every $R=50$ epochs of training
(Alg.~\ref{algo:OMLAadv}). We start with 1,000 training data samples considering a 9:1 split for training and validation. We
augment 200 adversarial samples at each \ac{SA} iteration. We trained for 350 epochs in total.

\subsection{Comparing Attack Models}\label{sec:comparingOMLAmodels}

\begin{table*}[!htb]
\caption{Predicted Attack Accuracy (\%) for Different Adversarial Models}
\footnotesize
\label{tab:kpaUsingALMOST}
\setlength\tabcolsep{3pt}
\resizebox{2\columnwidth}{!}{
\begin{tabular}{lcrrrrrrrrrrrrrr}
\toprule
\multirow{3}{*}{Variant} & \multirow{3}{*}{Key-size} & \multicolumn{14}{c}{Benchmarks} \\ \cmidrule(lr){3-16}
 & & \multicolumn{2}{c}{c1355} & \multicolumn{2}{c}{c1908} & \multicolumn{2}{c}{c2670} & \multicolumn{2}{c}{c3540} & \multicolumn{2}{c}{c5315} & \multicolumn{2}{c}{c6288} & \multicolumn{2}{c}{c7552} \\ \cmidrule(lr){3-4} \cmidrule(lr){5-6} \cmidrule(lr){7-8} \cmidrule(lr){9-10} \cmidrule(lr){11-12} \cmidrule(lr){13-14} \cmidrule(lr){15-16}
 & & \multicolumn{1}{c}{resyn2} & \multicolumn{1}{c}{random} & \multicolumn{1}{c}{resyn2} & \multicolumn{1}{c}{random} & \multicolumn{1}{c}{resyn2} & \multicolumn{1}{c}{random} & \multicolumn{1}{c}{resyn2} & \multicolumn{1}{c}{random} & \multicolumn{1}{c}{resyn2} & \multicolumn{1}{c}{random} & \multicolumn{1}{c}{resyn2} & \multicolumn{1}{c}{random} & \multicolumn{1}{c}{resyn2} & \multicolumn{1}{c}{random} \\
 \midrule
\multirow{2}{*}{$M_A^{resyn2}$}  & 64   &  57.52  & 54.21  & 59.01 &  50.57 &  58.00& 51.17 & 59.63 & 52.26& 62.62  & 57.46 & 52.21  &  53.32  & 66.33 & 58.46  \\
& 128   & 59.36   & 53.21  & 62.12 & 57.56  & 59.26 & 52.32 & 60.25 & 53.21 & 68.95  & 58.55    & 53.31  & 51.26   & 71.21 & 59.89 \\
\midrule
\multirow{2}{*}{$M_A^{random}$}  & 64   &  63.71  & 59.97  & 51.63 &  53.36 & 61.00 & 57.78 & 62.38 & 58.84 & 59.96 & 59.93 & 56.63  & 53.32& 66.33  & 60.17 \\
& 128    & 54.58 & 55.67  & 57.85 & 58.55 & 62.21 & 57.85&  54.20 & 55.87    &  65.41  & 61.25 & 52.26  & 51.88   & 65.25 & 61.56 \\
\midrule
\multirow{2}{*}{$M_A^*$}  & 64   &  61.94  & 61.05  & 53.27 & 55.55  & 63.00 & 62.50 & 62.38 & 61.37& 61.61  & 61.35    & 53.09  & 54.14   & 63.36 & 63.18 \\
& 128   &  59.36  & 58.89  & 61.05 & 62.10  & 61.89 & 63.45 & 59.98 & 63.42 & 67.88  &  69.25   & 52.99  & 54.58   & 69.25 &  66.59\\
\bottomrule
\end{tabular}
}
\smallerspacebelowfigure
\end{table*}

Here, we first analyze the prediction accuracy of various attack models when attacking the original locked circuit synthesized using resyn2 ($T_{resyn2}$) and in attacking the locked circuit synthesized with 1000 random synthesis recipes (the ``random set''). 
Table~\ref{tab:kpaUsingALMOST} shows the results, where the reported accuracy on the random set (\textit{random} column) is the average achieved accuracy. 

We observe a clear gap in accuracy when using $M_A^{resyn2}$ to attack $T_{resyn2}$ in comparison to its accuracy in attacking the random set. 
This shows that $M_A^{resyn2}$ learns the structural changes resulting from the application of the defender's synthesis recipe. 
However, the model accuracy suffers severely in attacking the random set. 
We observe that the accuracy reduced drastically $\sim1\%-9\%$ with an average of 4.8\%. 

In contrast, the accuracy of $M_A^{random}$ varies less in attacking $T_{resyn2}$ compared to attacking the random set. 
Also, the accuracy when attacking the random set is better than the accuracy exhibited by the $M_A^{resyn2}$ model. 

The $M_A^*$ model has more consistent accuracy when attacking $T_{resyn2}$ and the random set ($0.18\%-2.28\%$). 
$M_A^*$ consistently achieves higher accuracy than the other models on the random set. This indicates good generalization of $M_A^*$, suggesting that it is a good fit as the accuracy evaluator in \solution{}'s black-box optimization.

\subsection{Generating \solution{} Synthesis Recipe}
\begin{figure*}[!ht]
\captionsetup[subfigure]{labelformat=empty}
    \centering
    \subfloat[]{\includegraphics[width=0.55\columnwidth]{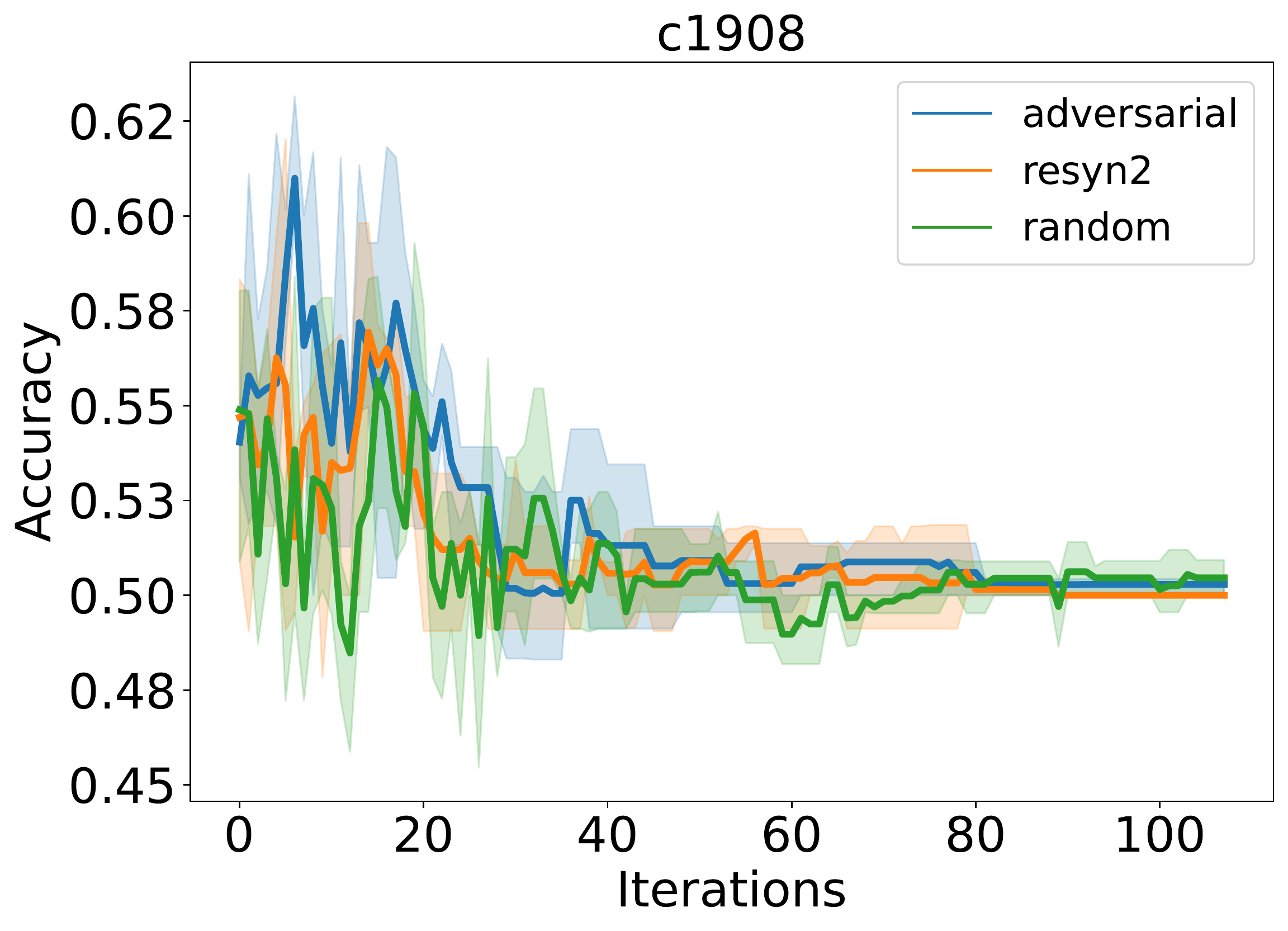}}\qquad
    \subfloat[]{\includegraphics[width=0.55\columnwidth]{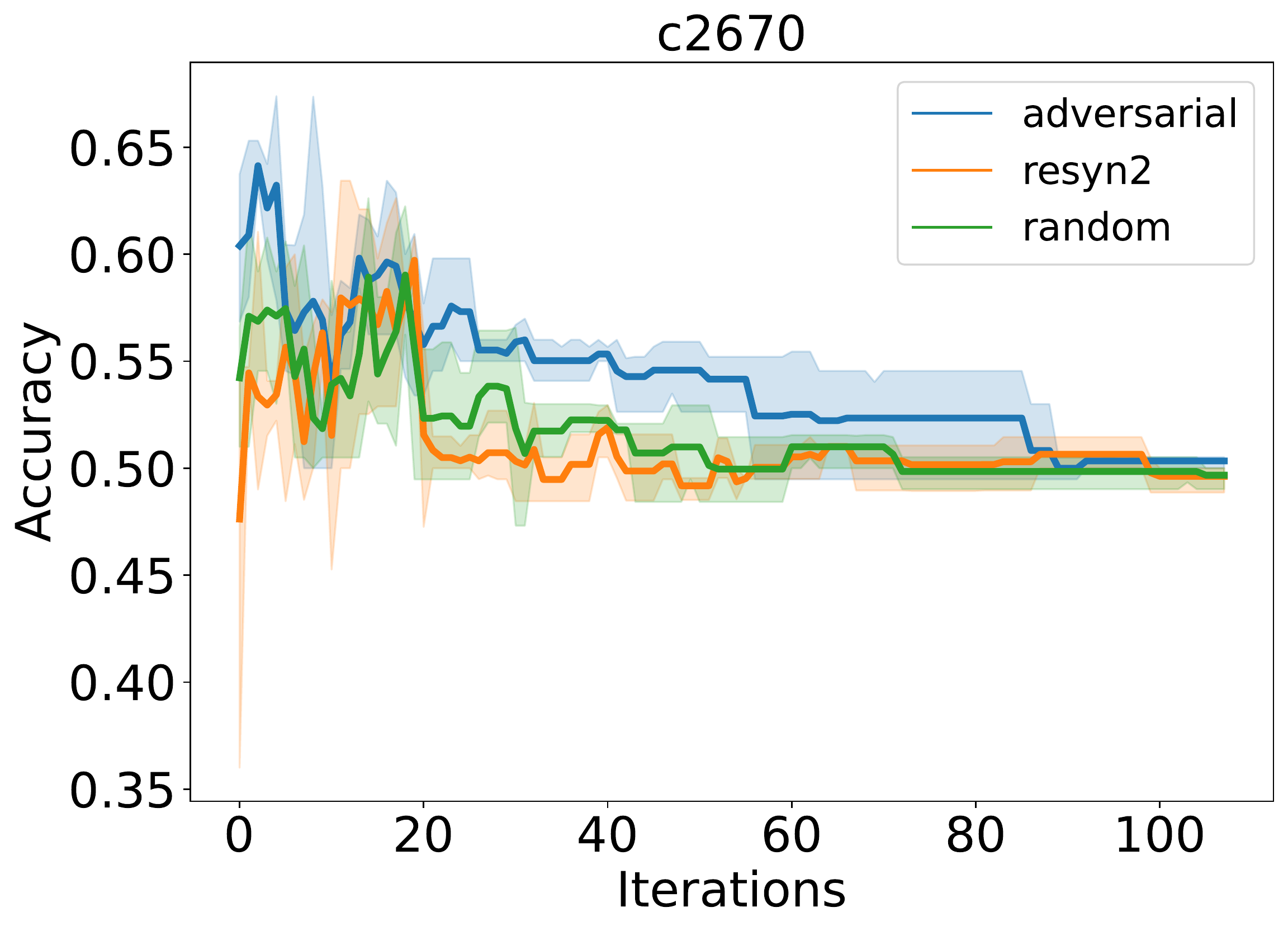}}\qquad
    \subfloat[]{\includegraphics[width=0.55\columnwidth]{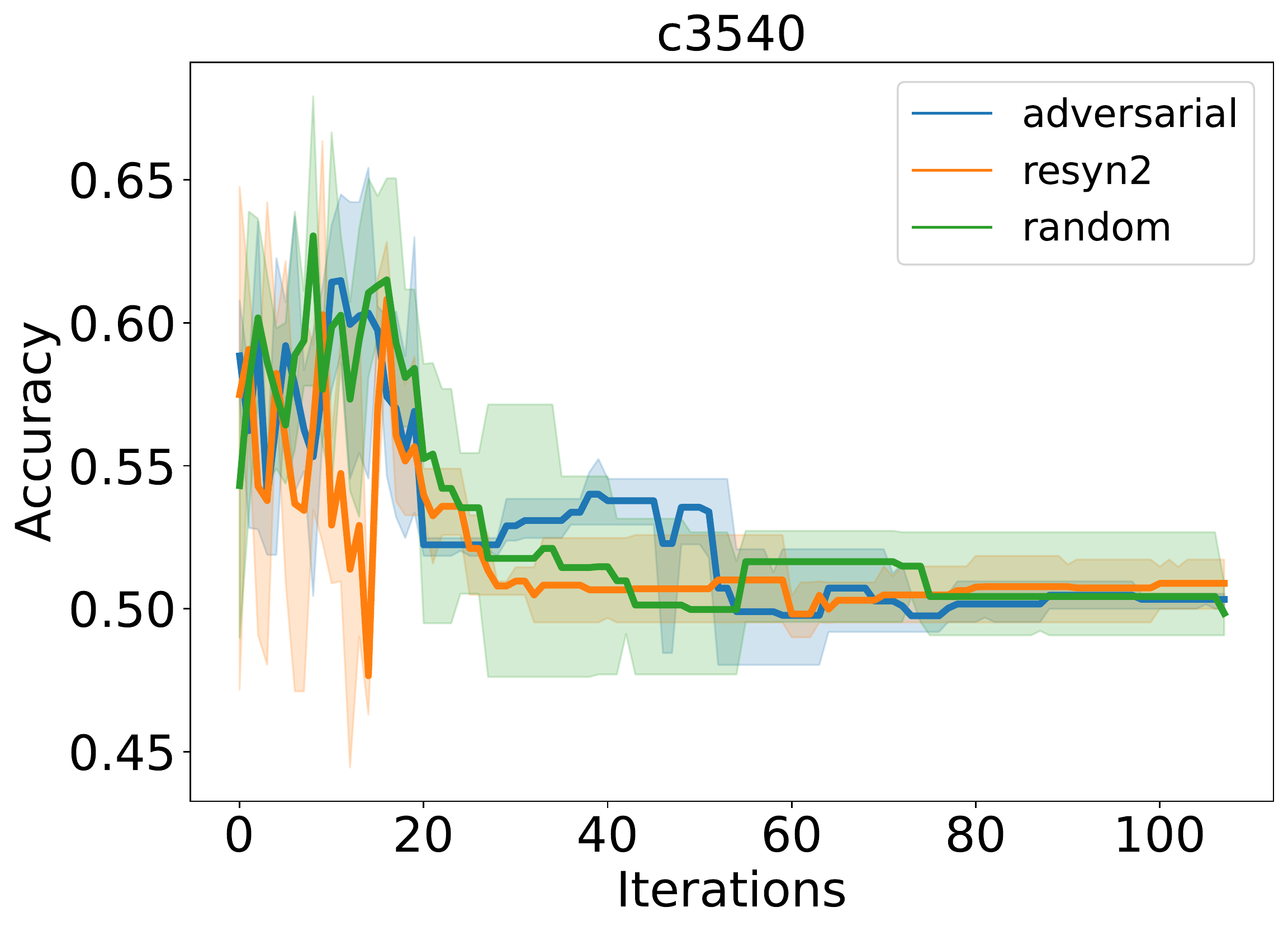}} \\ \smallerspacebelowfigure
    \subfloat[]{\includegraphics[width=0.55\columnwidth]{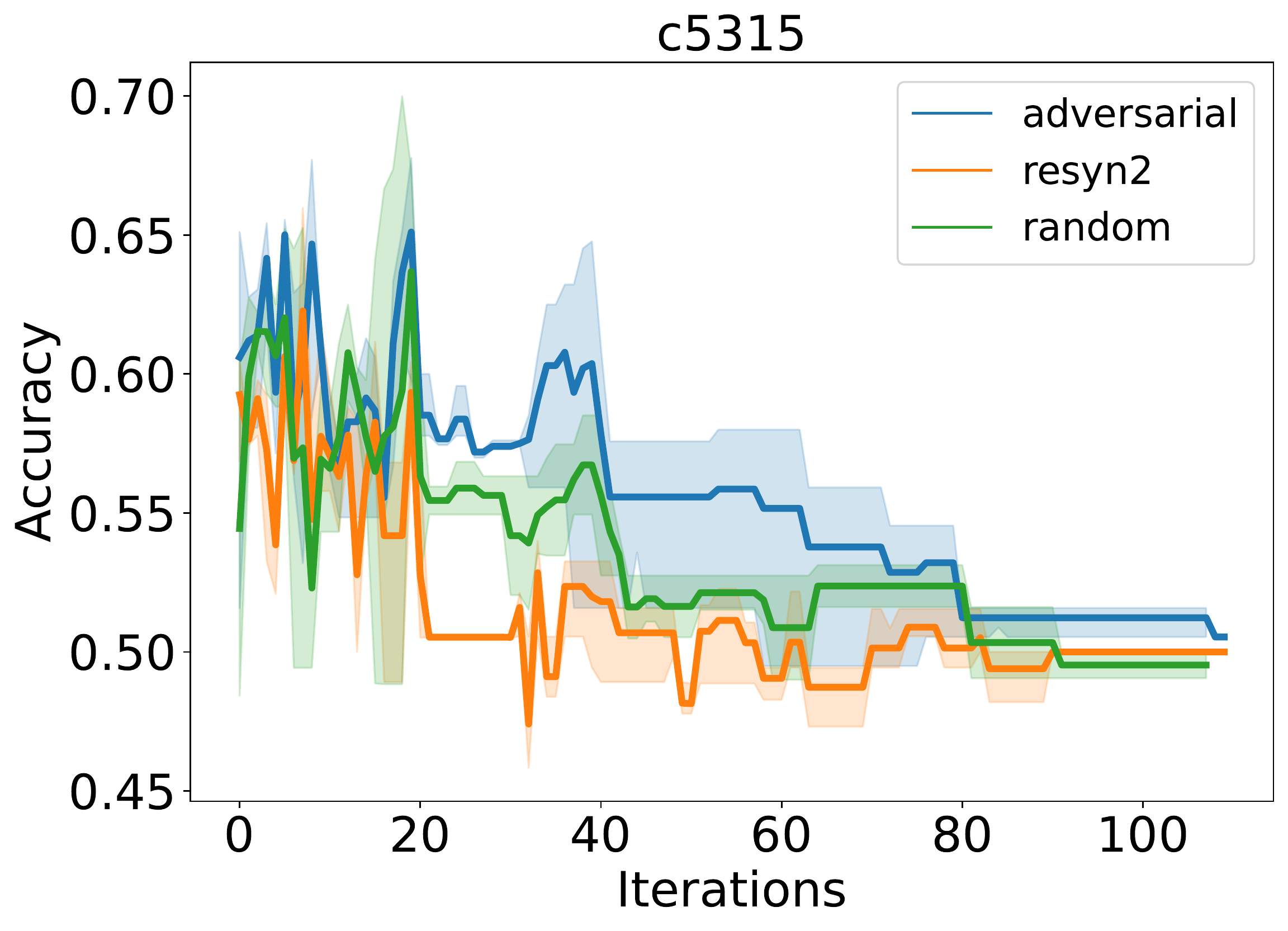}}\qquad
    \subfloat[]{\includegraphics[width=0.55\columnwidth]{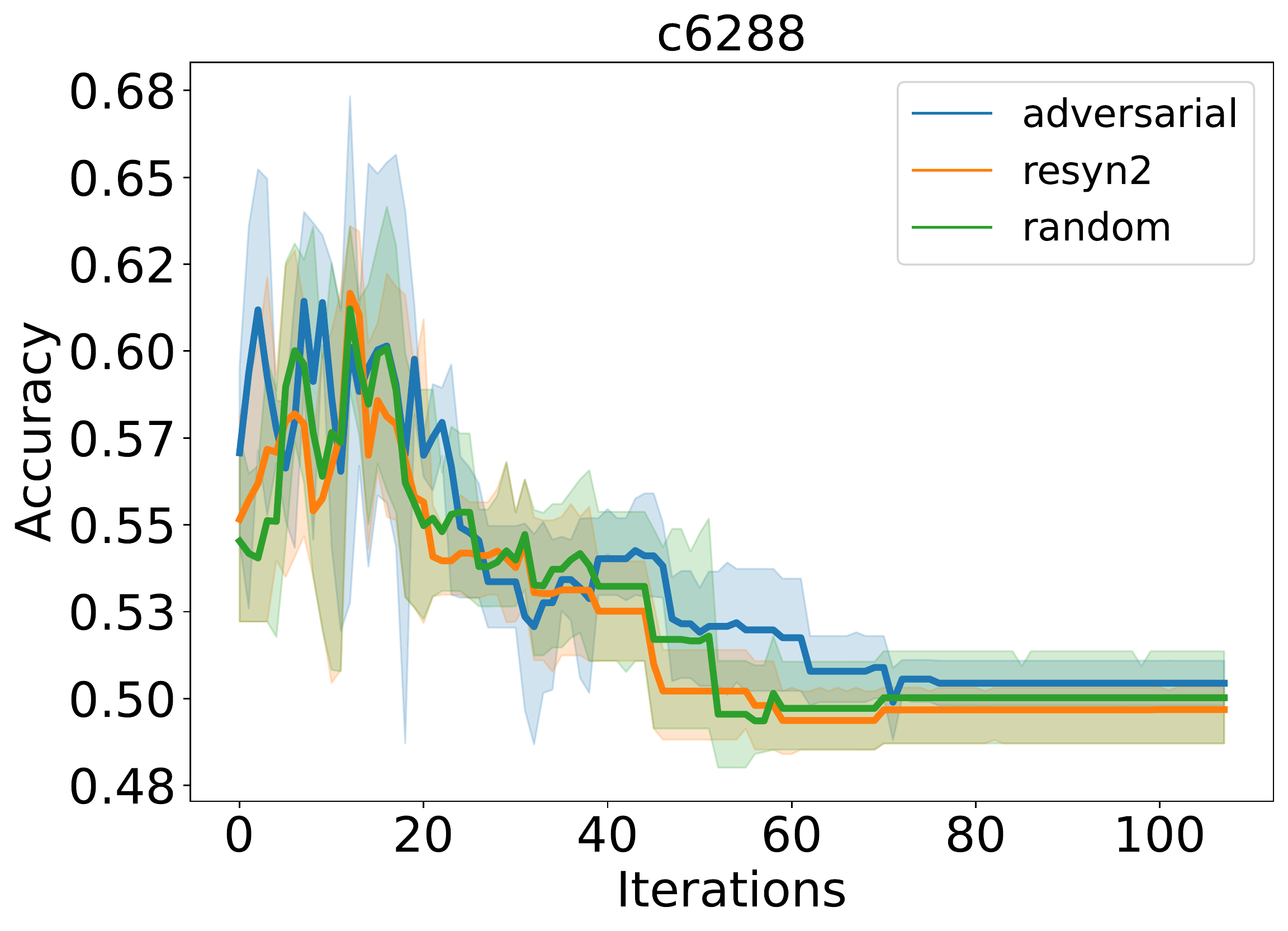}}\qquad
    \subfloat[]{\includegraphics[width=0.55\columnwidth]{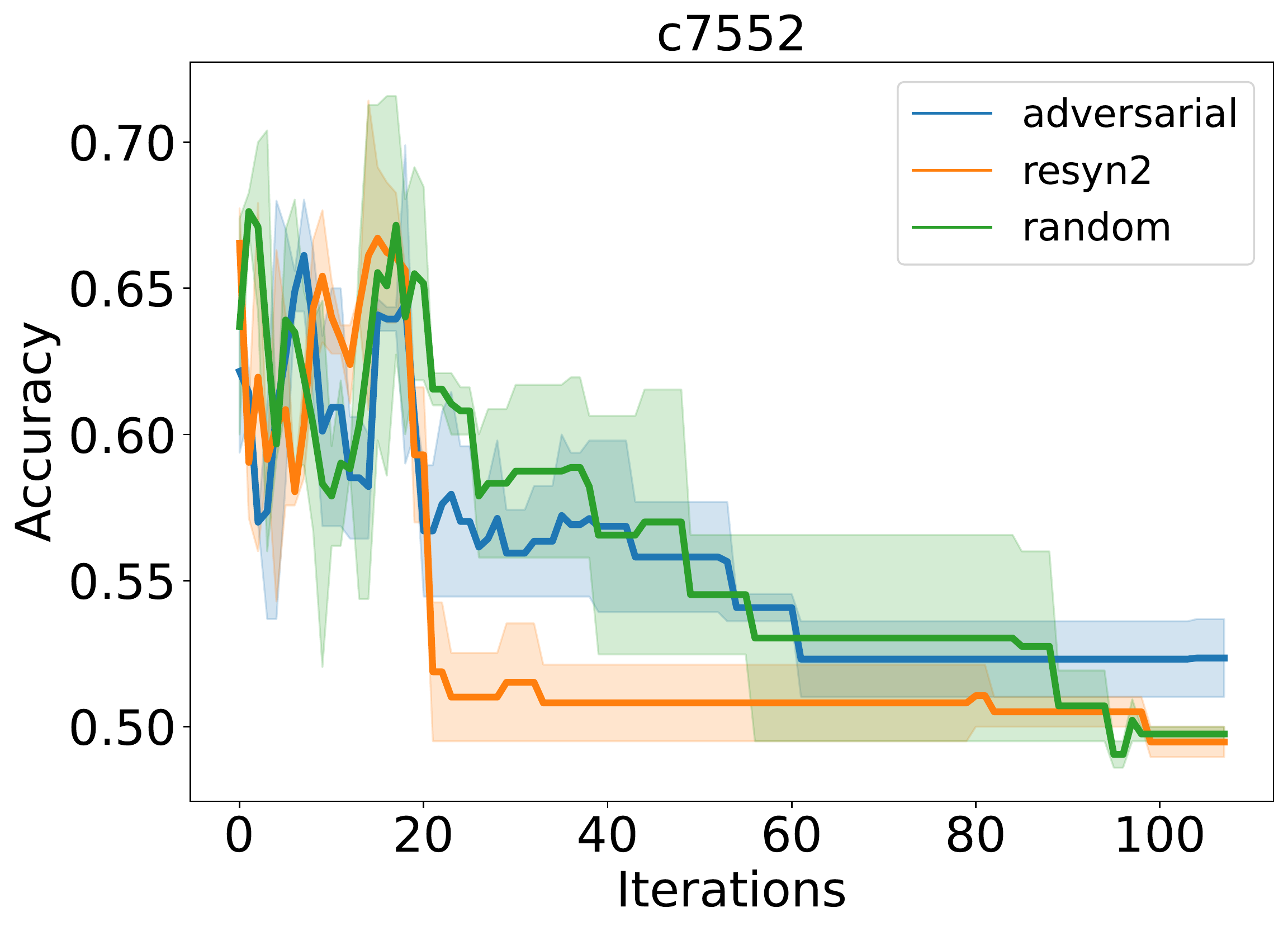}} 
    \caption{Simulated annealing-based recipe search for minimizing attack accuracy to 50\% or 0.5. \blue{$M_A^*$} model consistently has better attack accuracy and thus SA takes more iterations to find a recipe resulting in accuracy $\sim 50\%$.}
    \label{fig:almostSimulatedAnnealing}
    \smallerspacebelowfigure
\end{figure*}


We generate $S_{ALMOST}$ using the \ac{SA}-based recipe generator such that the attack accuracy using $M_A^{ALMOST}$ is $\sim$ 50\%. 
We run \ac{SA} for 100 iterations using an initial temperature of 120 and \texttt{acceptance=1.8}. Fig.~\ref{fig:almostSimulatedAnnealing} illustrates the \ac{SA}-based recipe generation on ISCAS benchmarks. For showing the effectiveness of \solution{} using the $M_A^*$ model, we compare our results to two other evaluators: $M_A^{resyn2}$ and $M_A^{random}$. \textcolor{blue}{Blue} line represents the attack accuracy estimated by $M_A^*$ on the locked netlist synthesized with the recipe generated during simulated annealing. \textcolor{orange}{Orange} and \textcolor{teal}{green} represent the attack accuracy trend for $M_A^{resyn2}$ and $M_A^{random}$, respectively. 

There is a consistent trend in all plots: \ac{SA} search using $M_A^*$ requires more iterations to find a synthesis recipe where accuracy goes to $\sim 50\%$; with a clear pattern on benchmarks c2670, c3540, c5315, and c7552. 
This follows our intuition: $M_A^*$ is trained with enough subgraph neighborhood diversity (netlist localities with key-gates 0 and 1). 
Hence, the \ac{SA} recipe generator requires more time to find a synthesis recipe in the search space that is unfamiliar to $M_A^*$. 
In contrast, using $M_A^{resyn2}$, \ac{SA} will quickly find a synthesis recipe where the accuracy falls to $\sim 50\%$. 
However, it is possible that there is a substantial accuracy gap between $M_A^{resyn2}$ and $M_A^{S}$ models for a particular recipe $S$ generated by \ac{SA}-based recipe generator which can give a false indication that the recipe will lead to general ML attack resilience. 
$M_A^{random}$ performed similarly to $M_A^*$, however, there is a wide variation in accuracy obtained by $M_A^{random}$ making it not a great choice for the defender's proxy model of an attacker. For c2670, c5315, and c7552 where accuracy could not reach $\sim 50\%$ within the budgeted iterations, we pick the synthesis recipe obtained at the end.


\subsection{Comparing \solution{} Synthesis Recipe with Resyn2}

\begin{table}[!htb]
\caption{Attack Accuracy (\%) Considering SOTA Attacks}
\footnotesize
\label{tab:kpaUsingALMOST_comparison}
\setlength\tabcolsep{2.5pt}
\resizebox{\columnwidth}{!}{
\begin{tabular}{cccrrrrrrr}
\toprule
\multirow{2}{*}{Attack} & \multirow{2}{*}{Keybits} & \multirow{2}{*}{Recipe} & \multicolumn{7}{c}{Benchmarks}  \\ \cmidrule(lr){4-10}
 & &   & \multicolumn{1}{c}{c1355} & \multicolumn{1}{c}{c1908} & \multicolumn{1}{c}{c2670} & \multicolumn{1}{c}{c3540} & \multicolumn{1}{c}{c5315} & \multicolumn{1}{c}{c6288} & \multicolumn{1}{c}{c7552} \\ \midrule
\multirow{4}{*}{OMLA} & \multirow{2}{*}{64}  & resyn2  &  57.52   & 59.01  &    58.01 & 59.63   & 62.62    & 52.51 & 66.33   \\
 &  & \solution{}  &  \textbf{54.18}  &    \textbf{47.80} & \textbf{49.78}   &    \textbf{46.57} & \textbf{49.78}     & \textbf{49.88} & \textbf{55.55}   \\
 & \multirow{2}{*}{128}    & resyn2  &  59.36   & 62.12  & 59.26    &  60.25  &68.95 &53.31    & 72.21 \\
 & & \solution{}  & \textbf{51.87}   & \textbf{49.81}   & \textbf{52.11}   & \textbf{48.92}   & \textbf{52.33}     &\textbf{50.00} & \textbf{51.88}  \\ \cmidrule(lr){1-10}
\multirow{4}{*}{SCOPE} & \multirow{2}{*}{64}  & resyn2  & 60.94   & 51.56  & 35.94  & \textbf{34.38}  &  \textbf{45.31}   & 53.13&    \textbf{40.63}\\
 &  & \solution{}  &  \textbf{56.25}  &    \textbf{48.44} & \textbf{31.25}   &    37.50 & 57.81     & \textbf{51.56} & 43.75   \\ 
 & \multirow{2}{*}{128}    & resyn2  &  51.56   & 46.09  &  \textbf{29.68}   & 36.71   & \textbf{37.50}    & 59.37 & 46.09  \\
 & & \solution{}  &  \textbf{50.78}  & 46.09   & 35.15   & 36.71   & 39.06     & \textbf{53.91} & \textbf{45.31} \\ \cmidrule(lr){1-10}
\multirow{4}{*}{Redundancy} & \multirow{2}{*}{64}  & resyn2  & \textbf{32.81}    & 37.50 & \textbf{28.13}   & 50.00   & 50.00   & 34.38 & 35.94   \\
 &  & \solution{}  &  39.06  &  37.50 & 31.25   &  \textbf{45.31} & 50.00     & \textbf{31.25} & \textbf{32.81}   \\
 & \multirow{2}{*}{128}    & resyn2  &  39.84   & \textbf{35.93}  &  21.09 & \textbf{41.40}   & 41.40     & \textbf{31.25} & 37.50 \\
 & & \solution{}  & \textbf{35.15}   & 42.96   & \textbf{19.53}   & 44.53   & \textbf{39.84} & 34.38     & \textbf{35.16}  \\ 
\bottomrule
\end{tabular}
}
\smallerspacebelowfigure
\end{table}

After generating the recipes, we evaluate the efficacy of \solution{} synthesized circuits in thwarting the \ac{OMLA}.
Table~\ref{tab:kpaUsingALMOST_comparison} compares OMLA attacks on locked circuits synthesized using \textit{resyn2} versus \solution{} generated synthesis recipe. On most benchmarks, our test accuracy is $\sim 50\%$. There is a  $3\%-12\%$ drop in accuracy  which is substantial. The proposed approach, which uses a proxy model in the loop, found a synthesis recipe from the search space that can transform the netlist structure in a way that thwarts \ac{OMLA} from obtaining high accuracy.
We also report attack results from applying the redundancy attack~\cite{redundancy} and SCOPE~\cite{alaql2021scope}. \solution{} synthesized circuits are more resilient than resyn2 synthesized ones. 

\subsection{What Happens if the Attacker Re-Synthesizes?}

\begin{figure*}[!ht]
\captionsetup[subfigure]{labelformat=empty}
    \centering
    \subfloat[]{\includegraphics[width=0.48\columnwidth]{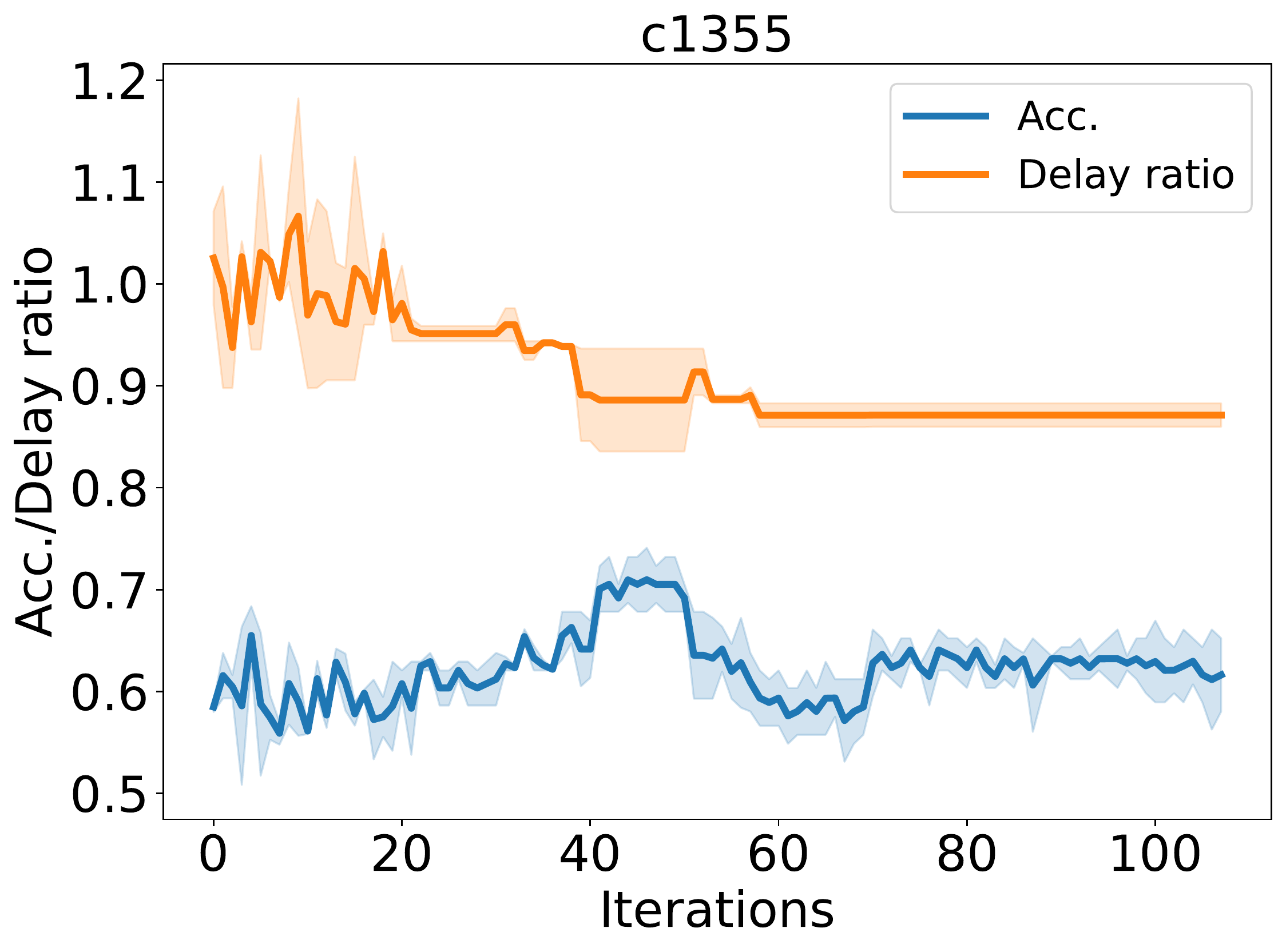}}\quad
    \subfloat[]{\includegraphics[width=0.48\columnwidth]{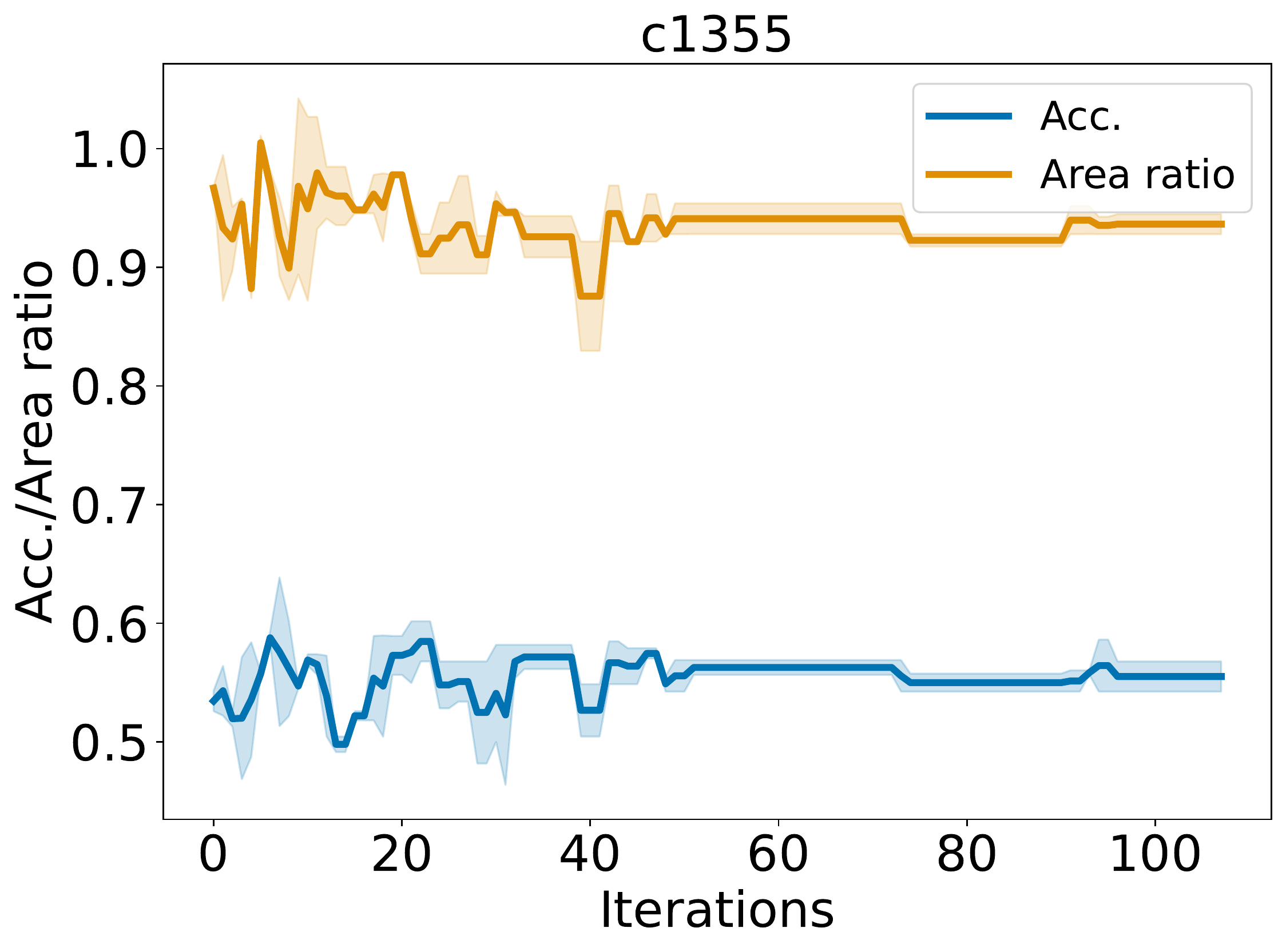}}\quad
    \subfloat[]{\includegraphics[width=0.48\columnwidth]{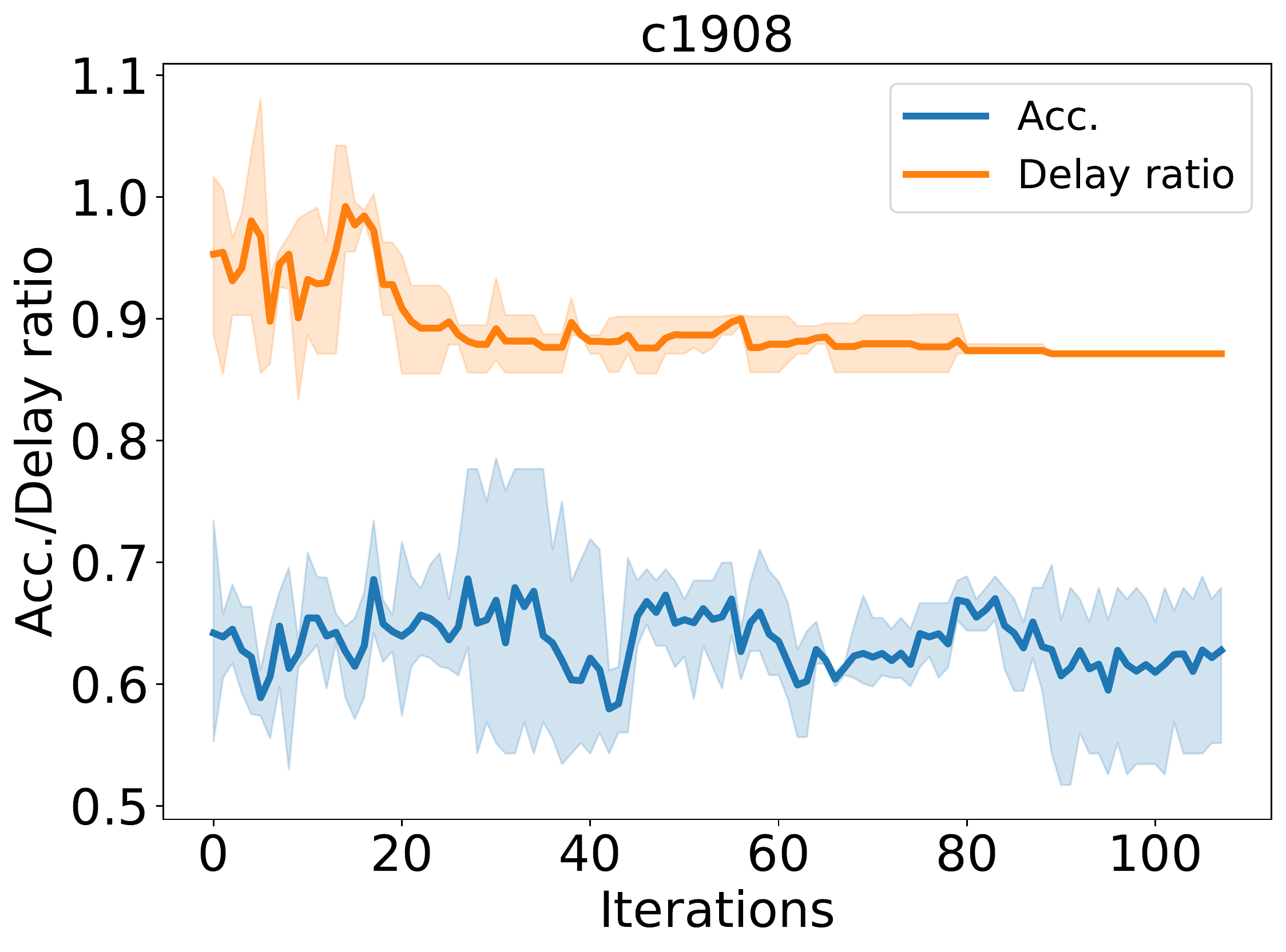}}\quad
    \subfloat[]{\includegraphics[width=0.48\columnwidth]{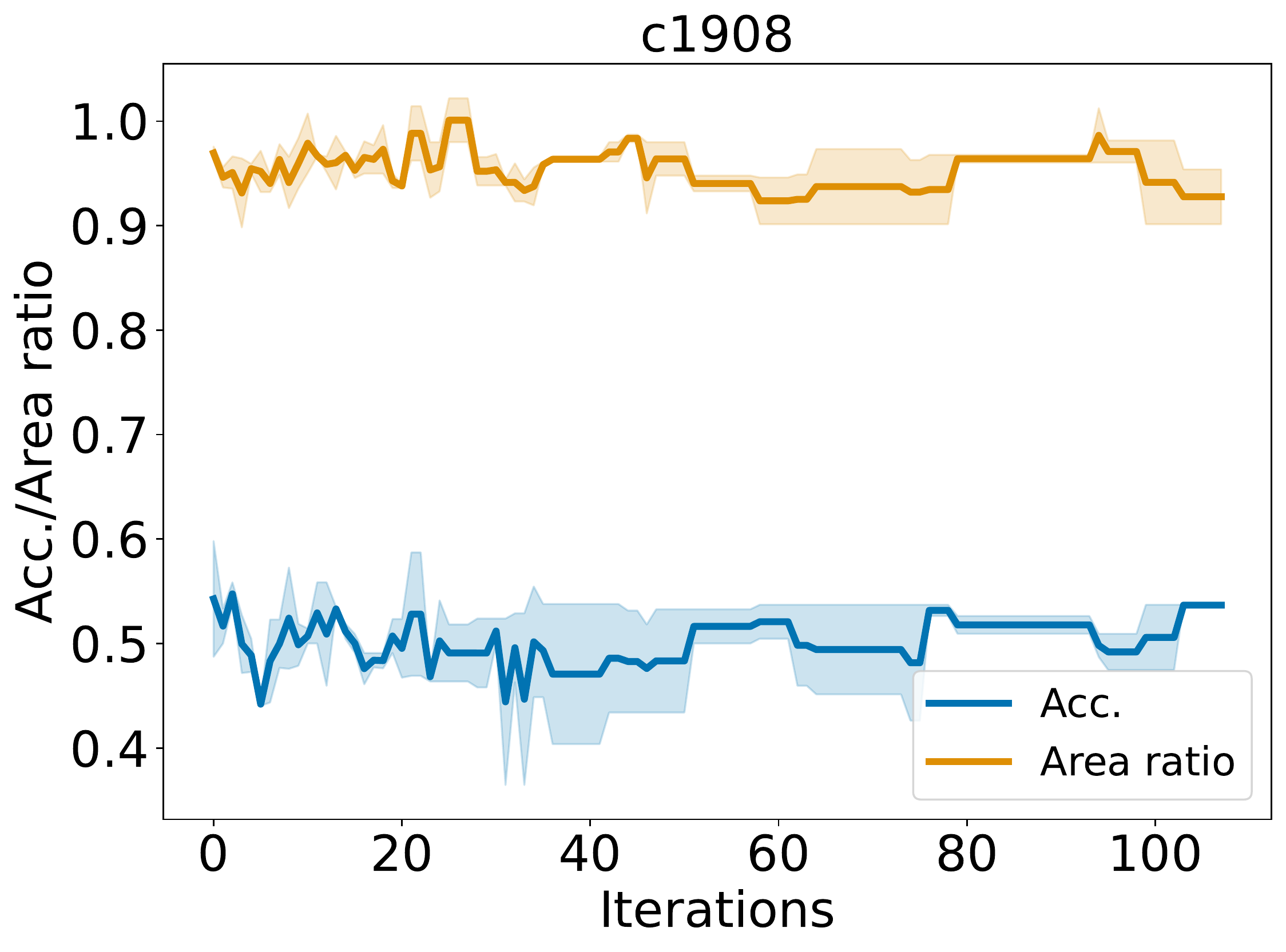}}\\ \smallerspacebelowfigure
    \subfloat[]{\includegraphics[width=0.48\columnwidth]{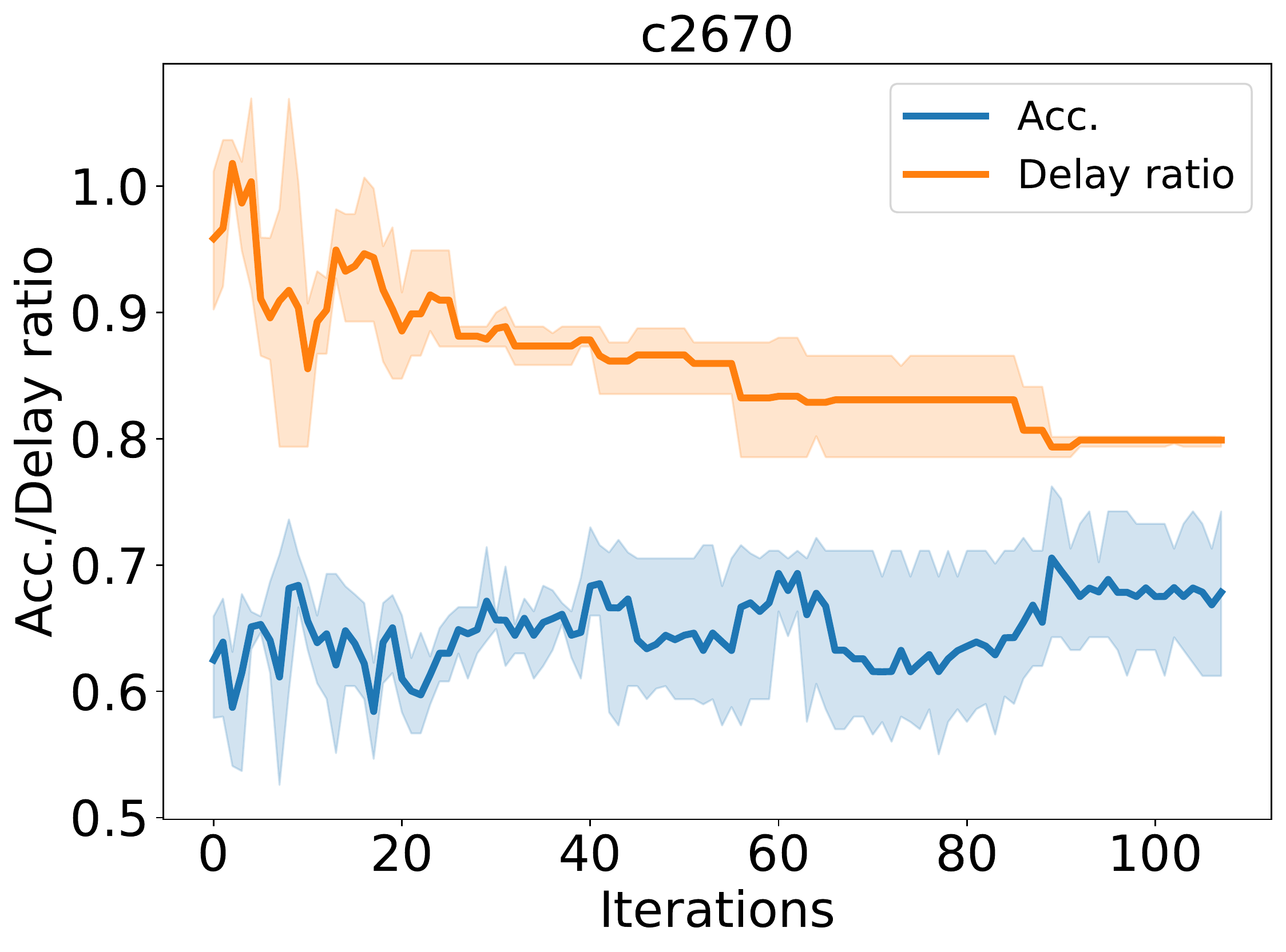}}\quad
    \subfloat[]{\includegraphics[width=0.48\columnwidth]{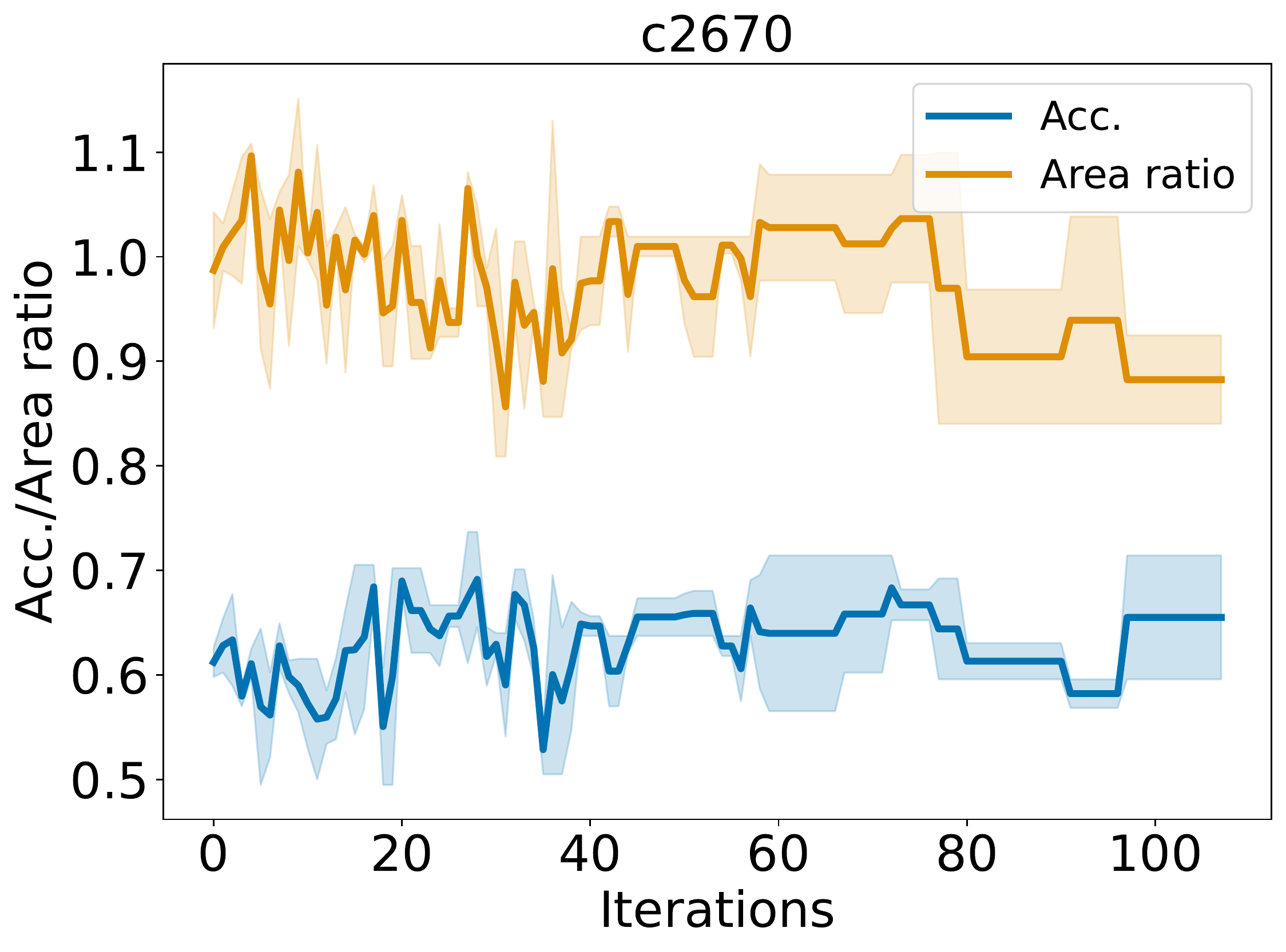}}\quad
    \subfloat[]{\includegraphics[width=0.48\columnwidth]{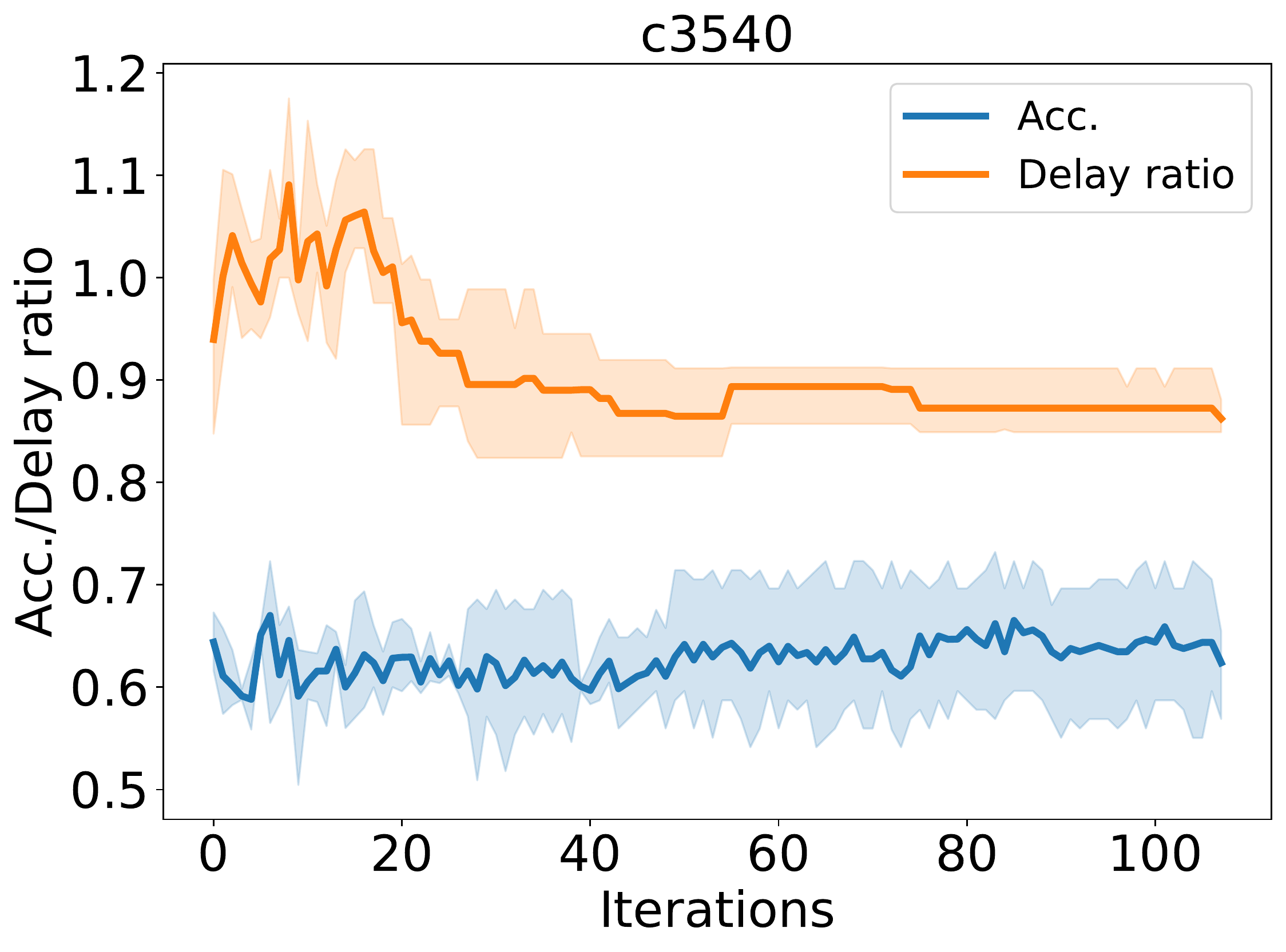}}\quad
    \subfloat[]{\includegraphics[width=0.48\columnwidth]{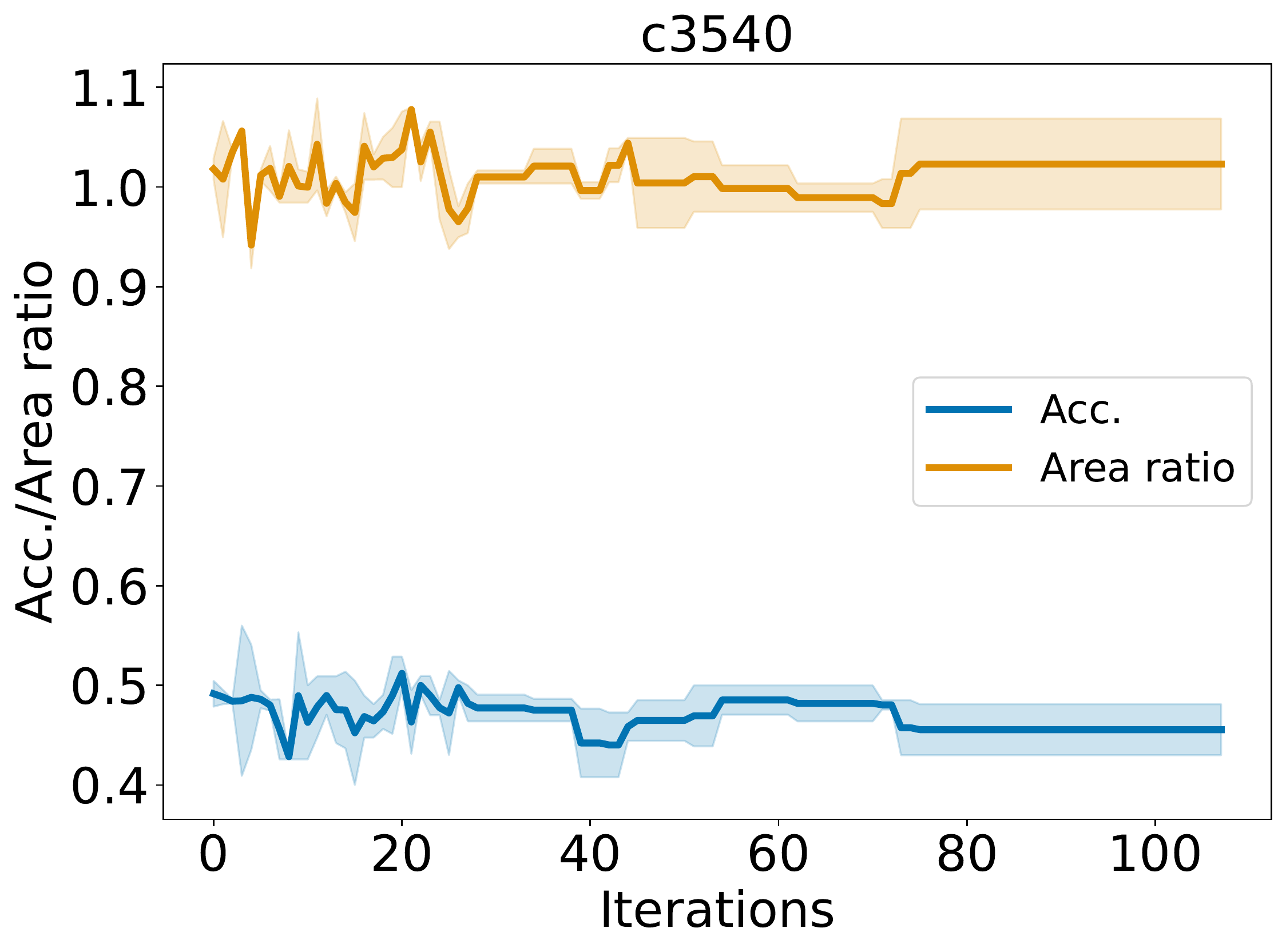}}\\ \smallerspacebelowfigure
    \subfloat[]{\includegraphics[width=0.48\columnwidth]{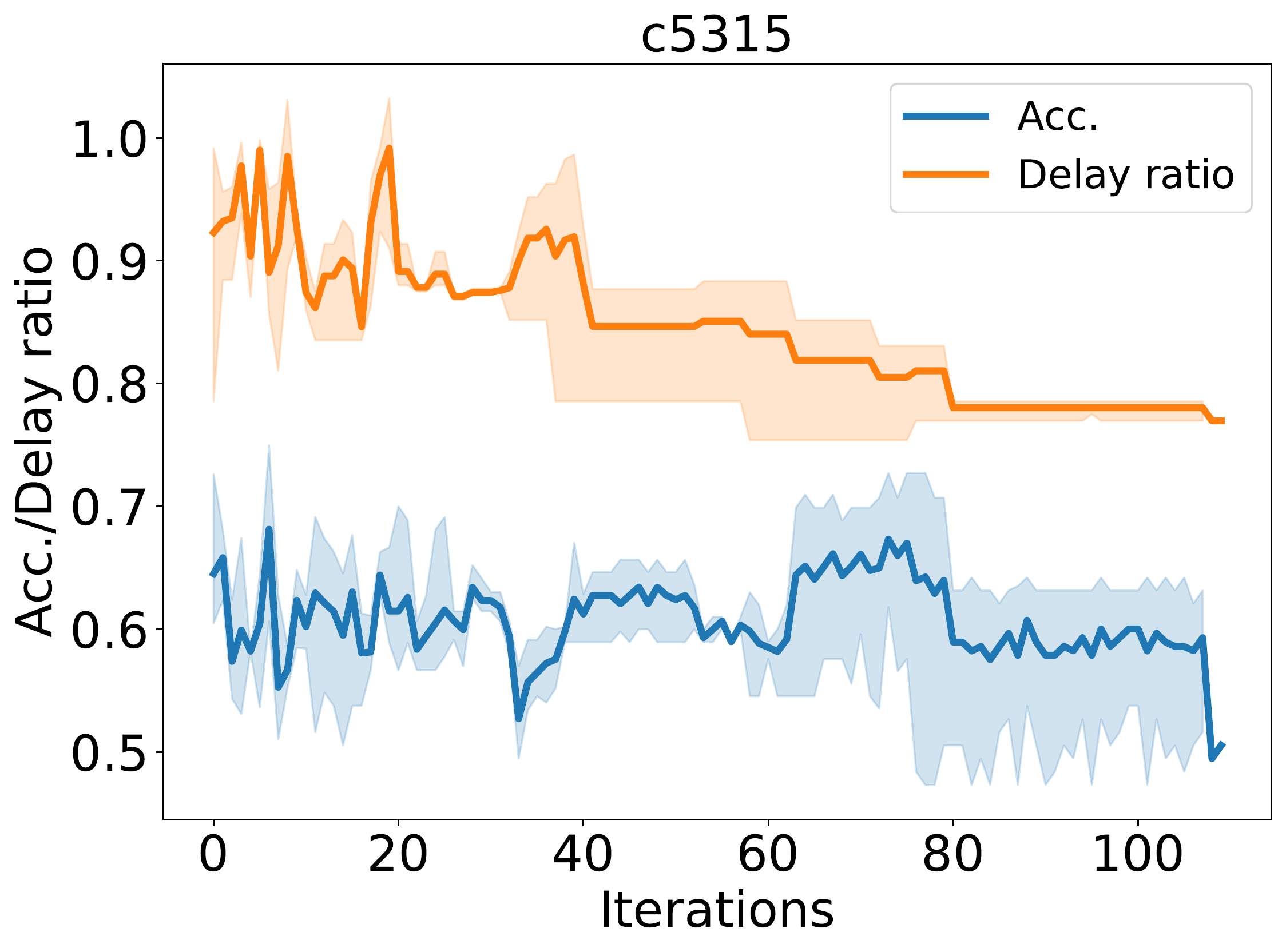}}\quad
    \subfloat[]{\includegraphics[width=0.48\columnwidth]{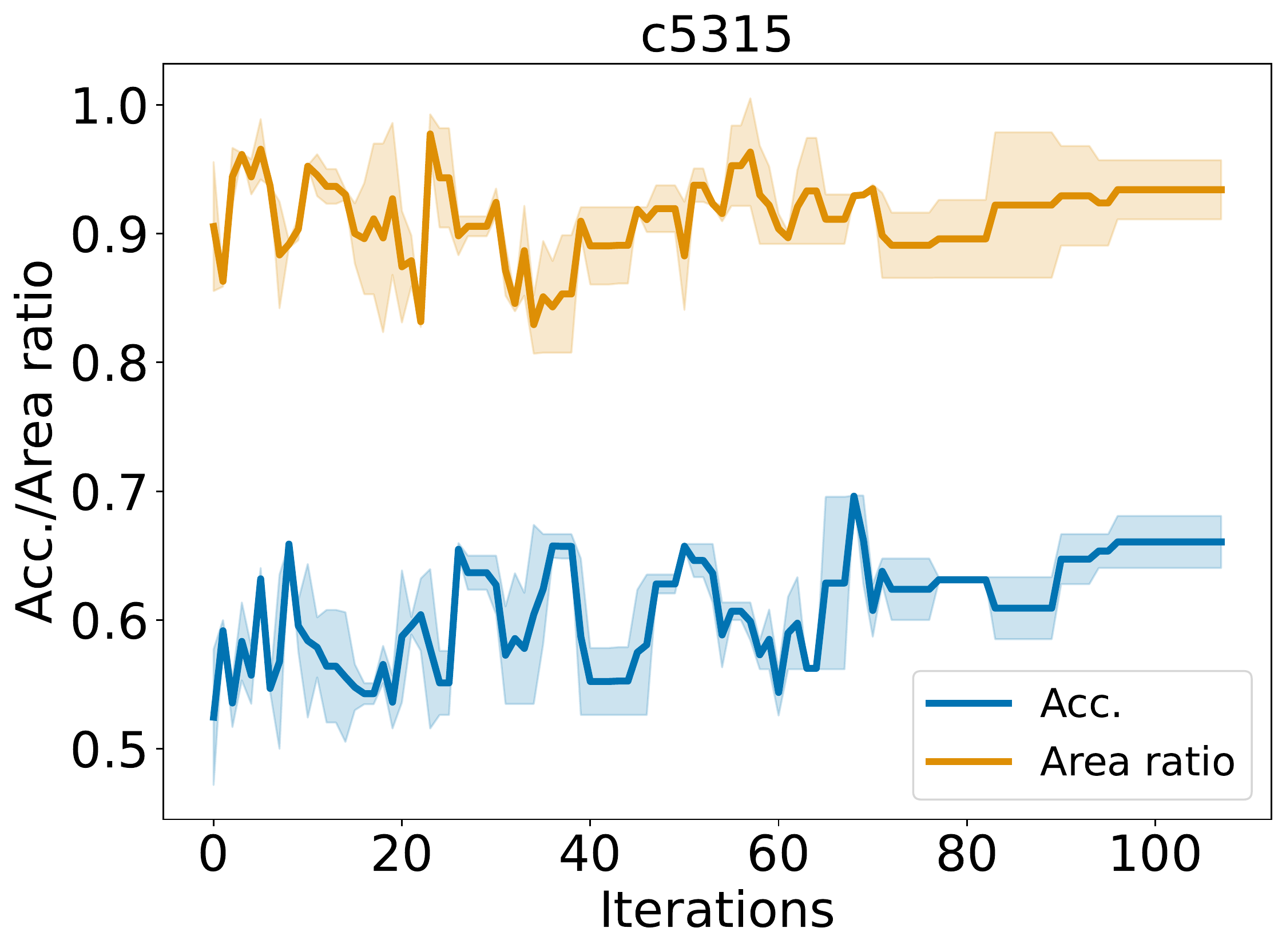}}\quad
    \subfloat[]{\includegraphics[width=0.48\columnwidth]{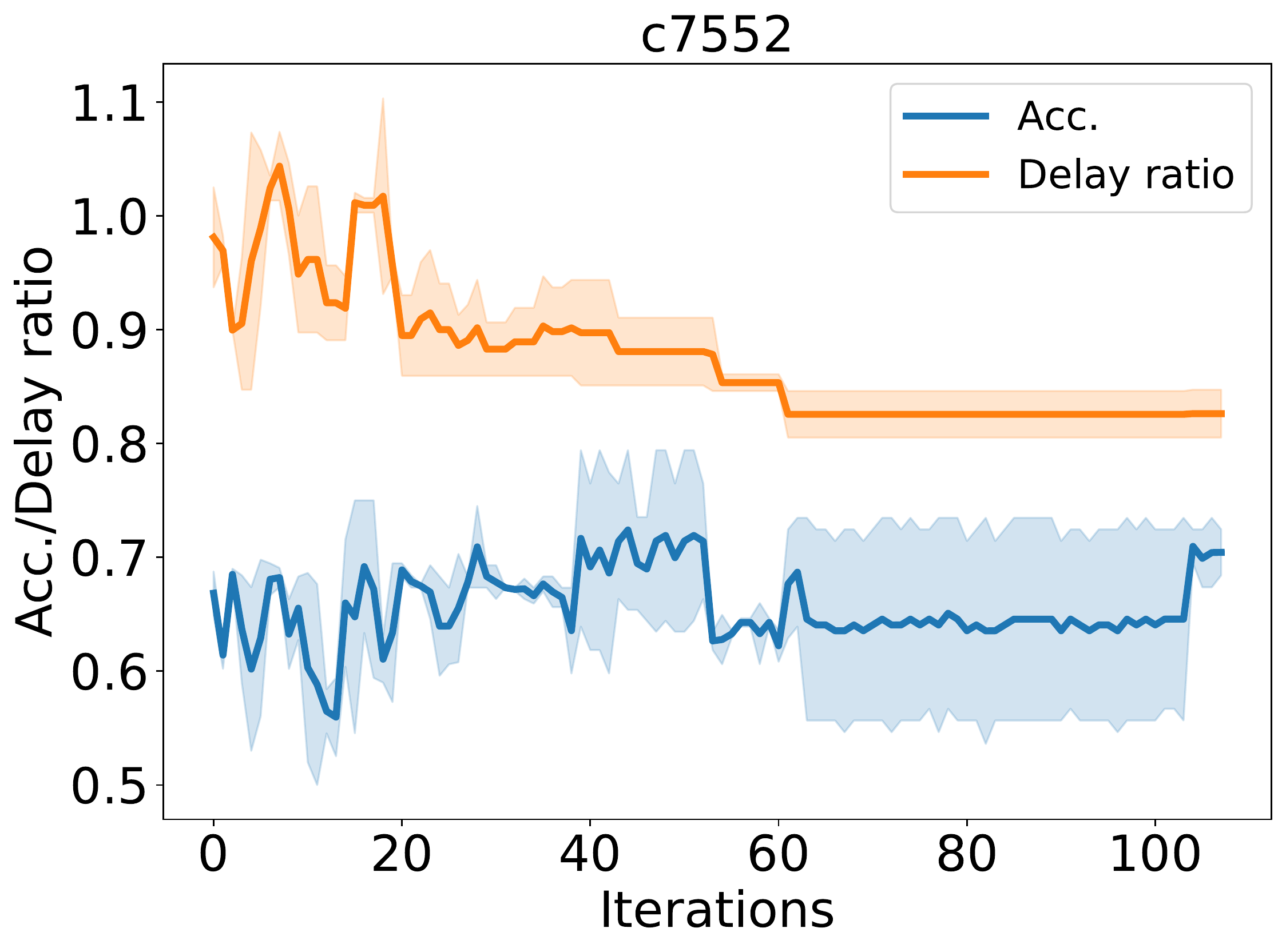}}\quad
    \subfloat[]{\includegraphics[width=0.48\columnwidth]{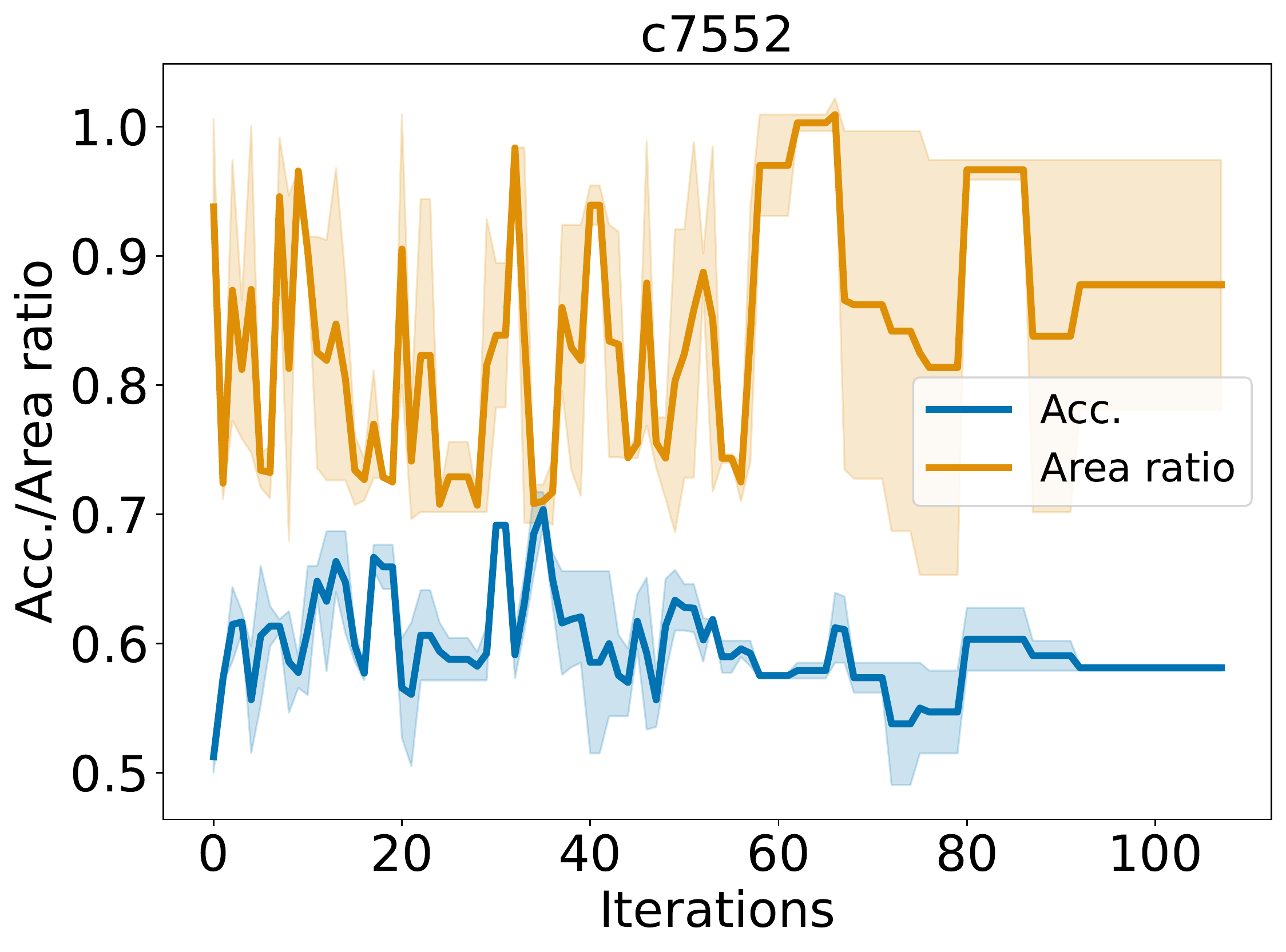}} 
    \caption{SA-based recipe search minimizing delay, area after ALMOST-driven synthesis. Note: attack accuracy does not correlate with delay, area optimization.}
    \label{fig:postDefenderSynthesis}
    
\end{figure*}

We assume the attacker can re-synthesize the \solution{} locked netlist with the aim to gain more information about how a design has been transformed. Hence, we run experiments to analyze whether an attacker can improve the accuracy of $M_A^{ALMOST}$ by re-synthesizing a locked netlist again and training $M_A^{ALMOST}$. We re-synthesize the \solution{} locked circuit for area optimization and delay optimization, assuming that this is the ``typical'' goal for synthesis. As such, a defender wants to avoid any 
correlation between area and/or delay optimization and attack accuracy since this can be exploited by the attacker. 
We use the \ac{SA}-based recipe generator with the \solution{} synthesized circuit as input and generate recipes targeting area/delay minimization. We take area and delay numbers of resyn2 as a baseline.

Fig.~\ref{fig:postDefenderSynthesis} shows the attack accuracy of $M_A^*$ on re-synthesized circuits and corresponding area/delay. \textcolor{blue}{Blue} denotes the attack accuracy of the \ac{ML} model and \textcolor{orange}{orange} denotes the normalized area/delay (compared with resyn2). Recipe length for re-synthesis is $L=10$ to match the length of resyn2. For delay minimization, \ac{SA} generates recipes minimizing delay. However, there is no noticeable variation in the area except for c2670 and c7552. There is no clear correlation between area/delay minimization and improvement/decline in attack accuracy. An attacker who re-synthesizes the circuit will not know which synthesis recipe to use to improve attack accuracy. 

\subsection{Analyzing power-performance-area (PPA) metrics}

\begin{table*}[!htb]
\caption{Power-performance-area (PPA) overhead (\%) for ALMOST synthesized circuits. -opt: No optimization, +opt: extreme optimization}
\centering
\footnotesize
\label{tab:ppaUsingALMOST}
\setlength\tabcolsep{3pt}

\begin{tabular}{@{}lcrrrrrrrrrrrrrr@{}}
\toprule
\multirow{3}{*}{Variant} & \multirow{3}{*}{Key-size} & \multicolumn{14}{c}{Benchmarks} \\ \cmidrule(lr){3-16}
 & & \multicolumn{2}{c}{c1355} & \multicolumn{2}{c}{c1908} & \multicolumn{2}{c}{c2670} & \multicolumn{2}{c}{c3540} & \multicolumn{2}{c}{c5315} & \multicolumn{2}{c}{c6288} & \multicolumn{2}{c}{c7552} \\ \cmidrule(lr){3-4} \cmidrule(lr){5-6} \cmidrule(lr){7-8} \cmidrule(lr){9-10} \cmidrule(lr){11-12} \cmidrule(lr){13-14} \cmidrule(lr){15-16}
 & & \multicolumn{1}{c}{-opt} & \multicolumn{1}{c}{+opt} & \multicolumn{1}{c}{-opt} & \multicolumn{1}{c}{+opt} & \multicolumn{1}{c}{-opt} & \multicolumn{1}{c}{+opt} & \multicolumn{1}{c}{-opt} & \multicolumn{1}{c}{+opt} & \multicolumn{1}{c}{-opt} & \multicolumn{1}{c}{+opt} & \multicolumn{1}{c}{-opt} & \multicolumn{1}{c}{+opt} & \multicolumn{1}{c}{-opt} & \multicolumn{1}{c}{+opt} \\
 \midrule
\multirow{2}{*}{Area}  & 64   &  +2.19  & +0.89  & -0.63 &  -0.95 &  -2.41 & -2.89  & +1.08 & +0.73 & +0.76  & +0.53 & +1.18  & +0.98  & +2.28 & +2.19  \\
& 128   & -0.05  & -0.65  & +2.32 & +1.98  & -0.38 & -0.57 & +0.94 & +0.67 & +0.04 & -0.05    & +3.08  & +2.79   & +0.84 & +0.76 \\
\midrule
\multirow{2}{*}{Delay} & 64 & -3.45   & -3.45  & -4.95  & -4.95 &  +18.31 & +18.31 & -0.46 & -0.46 & +5.00 & +3.75 & -0.93 & -0.70  & -15.24 & -15.24  \\
& 128    & +9.49 & +4.47  & -2.37 & -1.42 & +8.28 & +8.28 & +7.52 & +7.52 &  -2.69  & -2.69 & -6.70  & -6.49   & -7.01 & -7.01 \\
\midrule
\multirow{2}{*}{Power}  & 64   &  +3.36  & +2.25  & -0.28 &  -0.52 & -3.64 & -4.24 & +3.49 & +3.40 & -0.12 & -0.06 & -0.36  & -0.96 & +1.17  & +1.38 \\
& 128    & -1.10 & -1.37  & +2.20 & +2.05 & -0.17 & +0.12 &  -1.02  & -1.16    &  +0.81  & +0.63 & +2.57  & +2.28   & +0.81 & +0.48 \\
\bottomrule
\end{tabular}
\smallerspacebelowfigure
\end{table*}
To analyze the overhead \solution{} synthesized circuits bear for ML attack resilience, we present and analyze the PPA overhead of \solution{} synthesized designs by running Synopsys DC compiler in Table~\ref{tab:ppaUsingALMOST}. 
We consider two settings: (1) No optimization (-opt), and (2) Extreme optimization (+opt), where we enable ultra effort optimization along with the area recovery option. 
We use the PPA of the original locked netlist as a baseline. 
Area overhead varies in the range of $\sim \pm3\%$. Similarly, power overhead also varies in the range of $\sim \pm5\%$. For delay, circuits like c2670 have a relatively high overhead of around $18$\%. However, delay overhead is 15\% lower for c7552. On average, \solution{} generates \ac{ML} attack resilient circuits with low overhead.
\section{Conclusion and Future Work}
\solution{} mitigates oracle-less ML-based attacks on logic locking.  It uses synthesis tuning to make designs locked with a 100\% vulnerable locking approach attack resilient using suitable synthesis recioes, with low impacts on PPA metrics. It  applies to other locking techniques.  Future research directions include investigating the impact of synthesis transformations in creating indistinguishable key-gate localities for \ac{ML} attack resilient design, developing a generalized reinforcement learning-based synthesis engine to generate resilient designs, and jointly optimizing PPA and security metrics.




\section*{Acknowledgments}
This research was supported in part by NSF Award 2039607. The opinions, findings, and conclusions, or recommendations expressed are those of the author(s) and do not  reflect the views of any sponsors.



\bibliographystyle{IEEEtran}
\scriptsize{
    \bibliography{biblio.bib}
}


\end{document}